\documentclass[prl,reprint,superscriptaddress,preprintnumbers]{revtex4-1}
\pdfoutput=1
\usepackage{hyperref}
\usepackage[T1]{fontenc}

\usepackage{mathrsfs}

\usepackage{amsmath,amssymb,amsfonts,amsxtra,mathrsfs,graphics,graphicx,amsthm,epsfig,bm,longtable,float,color,tikz,mathtools,xfrac,footnote}
\restylefloat{table}
\usepackage{dsfont}
\usepackage{lipsum}
\usepackage{textgreek}
\usetikzlibrary{decorations.pathmorphing}
\usetikzlibrary{decorations.markings}
\usetikzlibrary{quotes,arrows.meta}
\usetikzlibrary{arrows,decorations.markings,calc,fadings,decorations.pathreplacing,patterns,decorations.pathmorphing,positioning}
\usepackage{tikz-cd}

\newcommand{\ii}{\mathrm{i}}
\newcommand{\?}{\;\!}

\newcommand{\be}{\begin{equation}} \newcommand{\ee}{\end{equation}}
\newcommand{\bea}{\begin{equation} \begin{aligned}} \newcommand{\eea}{\end{aligned} \end{equation}}

\def\U{\mathrm{U}}
\def\SO{\mathrm{SO}}

\def\SU{\mathrm{SU}}

\newcommand{\wt}{\widetilde}

\DeclareMathOperator{\Tr}{Tr}

\newcommand{\pd}{\partial}

\newcommand{\cI}{\mathcal{I}}

\newcommand{\cL}{\mathcal{L}}

\newcommand{\cN}{\mathcal{N}}
\newcommand{\cO}{\mathcal{O}}

\newcommand{\cS}{\mathcal{S}}

\newcommand{\cZ}{\mathcal{Z}}

\newcommand{\fg}{\mathfrak{g}}

\newcommand{\fs}{\mathfrak{s}}
\newcommand{\ft}{\mathfrak{t}}

\begin{document}

\title{Universal AdS black holes in theories with sixteen supercharges and their microstates}

\date{\today}

\author{Seyed Morteza Hosseini}
\affiliation{Kavli IPMU (WPI), UTIAS, The University of Tokyo, Kashiwa, Chiba 277-8583, Japan}
\email{morteza.hosseini@ipmu.jp}

\author{Alberto Zaffaroni}
\affiliation{Dipartimento di Fisica, Universit\`a di Milano-Bicocca, I-20126 Milano, Italy}
\affiliation{INFN, sezione di Milano-Bicocca, I-20126 Milano, Italy}

\email{alberto.zaffaroni@mib.infn.it}

\begin{abstract}

We provide a universal microscopic counting for the microstates of the asymptotically AdS black holes and black strings that arise as solutions of the half-maximal gauged supergravity in four and five dimensions. These solutions can be embedded in all M-theory and type II string backgrounds with an AdS vacuum and sixteen supercharges and provide an infinite set of examples dual to $\mathcal{N}=2$ and $\mathcal{N}=4$ conformal field theories in four and three dimensions, respectively. The counting is universal and it is performed by either studying the large $N$ limit of the relevant supersymmetric index of the dual field theory or by using the charged Cardy formula.

\end{abstract}



\maketitle

{\it Introduction}.--- The microscopic counting of black hole microstates is a fundamental question for all theories of quantum gravity. String theory has provided a microscopic explanation for the entropy of a class of asymptotically flat black holes \cite{Strominger:1996sh}. Despite the AdS/CFT correspondence \cite{Maldacena:1997re}, the analogous question for asymptotically anti de Sitter (AdS) black holes has remained elusive until recently, except for AdS$_3$. In the last few years there has been some progress, first for a class of magnetically charged supersymmetric black holes in AdS$_4$ \cite{Benini:2015eyy}, and later for a class of supersymmetric Kerr-Newman (KN) black holes in AdS$_5$ \cite{Benini:2018ywd,Choi:2018hmj}.  The AdS/CFT correspondence  provides a non-perturbative definition of quantum gravity in asymptotically AdS space in terms of a dual boundary quantum field theory (QFT) and  the black hole microstates appear as particular states in the boundary description. The entropy of a supersymmetric black hole with angular momentum $J$ and a set of conserved electric and magnetic charges is reproduced by counting the states with spin $J$ and the same quantum numbers in the dual QFT. Computations  in a strongly coupled supersymmetric QFT are difficult but the numbers of states of interest can be extracted from supersymmetric indices that  can be often evaluated using  exact non-perturbative techniques. Supersymmetric localization \cite{Pestun:2016zxk}, for example, allows one to reduce  the indices  to matrix models that can be evaluated in a saddle point approximation
when the number of colors $N$ is large,  which is the regime where  holography  applies.

Unfortunately,  asymptotically AdS black holes are difficult to find and not so many four- and five-dimensional examples are known besides those that can be embedded in AdS$_4\times S^7$ or AdS$_5\times S^5$ and some universal examples that arise from embedding minimal gauged supergravity into string compactifications \cite{Azzurli:2017kxo,Cabo-Bizet:2019osg,Kim:2019yrz,Benini:2020gjh}.  Some progress has been  made instead in constructing infinite classes of supersymmetric AdS$_2$ and AdS$_3$ solutions that can  arise as the near horizon limit of AdS black objects \cite{Couzens:2018wnk,Gauntlett:2019pqg}.  In this Letter, in the spirit of the above-mentioned universal examples, 
we consider a large class of black holes and black strings that arise as solutions of the half-maximal supergravity in AdS$_4$ and AdS$_5$. Such solutions  can be embedded in all AdS$_4$ and AdS$_5$ type II or M-theory backgrounds with sixteen supercharges. Indeed, for any supersymmetric solution of ten- or eleven-dimensional supergravity of the warped product form AdS$_D \times_w  M$, there is a consistent truncation to pure gauged supergravity in $D$ dimensions containing that solution and having the same supersymmetry \cite{Gauntlett:2007ma,Gauntlett:2007sm,Cassani:2019vcl,Malek:2017njj}. This observation calls for a {\it universal} large $N$ formula for the number of supersymmetric states  in conformal field theories (CFT) with eight supercharges (sixteen including conformal supersymmetries). We will indeed perform a universal counting of states in  CFTs with eight supercharges using the superconformal index and the charged Cardy formula. Extrapolating from known results 
about the large $N$ behaviour of the index in various limits, we precisely reproduce the Bekenstein-Hawking entropy of the black objects. We will check the results for the known classes of CFTs with eight supercharges that admit a holographic dual. We will  also provide a conjecture for the R-symmetry charge dependence of the $S^3$ free energy of 3d CFTs with eight supercharges. 
A similar universal computation was done in \cite{Azzurli:2017kxo} and in \cite{Cabo-Bizet:2019osg,Kim:2019yrz,Benini:2020gjh} for theories with four supercharges.  
For other related universal results see \cite{Benini:2015bwz,Bobev:2017uzs,Bobev:2019zmz}.

{\it $\cN=4$ gauged supergravity in five-dimensions}.---%
The bosonic part of minimal 5d $\cN = 4$ gauged supergravity \cite{Romans:1985ps,Awada:1985ep} consists of
the metric $g_{\mu \nu}$, a $\U(1)_R$ gauge field $a_\mu$, an $\SU(2)_R$ Yang-Mills gauge field $A^I_{\mu}$, $I=1,2,3$,
two antisymmetric tensor fields $B_{\mu \nu}^\alpha$, $\alpha = 4,5$, and one real scalar $\phi$.
The fermionic components are four gravitini $\psi_{\mu i}$, $i=1,2,3,4$, and four spin-$1/2$ fermions $\chi_i$.
These fields form the $\cN= 4$ gauged supergravity multiplet $(g_{\mu \nu}, \psi_{\mu i}, a_\mu, A_\mu^I, B_{\mu \nu}^\alpha, \chi_i, \phi)$.
The bosonic Lagrangian is given by
\bea
 \frac{1}{\sqrt{- g}} & \cL = - \frac{R}{4} + \frac{1}{2} (\partial^\mu \phi) (\partial_\mu \phi) + \frac{g_2}{8} \left( g_2 \xi^{-2} + 2 \sqrt{2} g_1 \xi \right) \\
 & - \frac{\xi^{-4}}{4} f^{\mu \nu} f_{\mu \nu} - \frac{\xi^2}{4} \Big( F^{\mu \nu I} F_{\mu \nu}^I + B^{\mu \nu \alpha} B_{\mu \nu}^\alpha \Big) \\
 & + \frac1{4\sqrt{- g}} \varepsilon^{\mu \nu \rho \sigma \tau} \left( \frac{1}{g_1} \varepsilon_{\alpha \beta} B^\alpha_{\mu \nu} D_{\rho} B^\beta_{\sigma \tau} - F^I_{\mu \nu} F^I_{\rho \sigma} a_\tau \right) ,\nonumber
\eea
where $\xi = \exp(\sqrt{2/3} \phi)$, and the field strengths are $f_{\mu} = \pd_{\nu} a_{\mu} - \pd_{\nu} a_{\mu}$, $F_{\mu \nu}^I =  \pd_{\nu} A_{\mu}^I - \pd_{\mu} A_{\nu}^I + g_2 \varepsilon^{IJK} A^J_\mu A_\nu^K$.
The theory with $g_2 = \sqrt{2} g_1 = 2 \sqrt{2}$ has an AdS vacuum with radius $\ell_{5} = 1$ that preserves all of the sixteen real supercharges.

{\it The universal  KN black hole in AdS$_5$}.--- There is a universal solution of $\cN=4$ gauged supergravity  corresponding to a supersymmetric, asymptotically AdS, black hole
with two electric charges $Q_1$ and $Q_2$ under $\U(1)^2\subset \U(1)_R\times \SU(2)_R$ and two angular momenta $J_1$ and $J_2$. It can be obtained as a particular case of the KN black holes
in AdS$_5\times S^5$ with angular momenta $(J_1,J_2)$ and electric charges $(Q_1,Q_2,Q_2)$ under the Cartan subgroup of $\SO(6)$ \cite{Gutowski:2004ez,Gutowski:2004yv,Chong:2005da,Chong:2005hr,Kunduri:2006ek}. The latter are solutions of the $\U(1)^3$ truncation of AdS$_5\times S^5$ and one can easily check that they become solutions  of the minimal 
$\cN=4$ gauged supergravity when two $\U(1)$ gauge fields are identified. The entropy can be compactly written as \cite{Kim:2006he}
\be
 \label{EntropyKNAdS5} S(J_I,Q_i) = 2 \pi \sqrt{Q_2^2+ 2 Q_1 Q_2 -2 a (J_1+J_2)} \, ,
\ee
where $a= \frac{\pi \ell_5^3}{8 G_{\rm N}^{(5)}}$, with $\ell_5$ being the radius of AdS$_5$ and $G_{\rm N}^{(5)}$ the Newton constant, is the central charge of the
dual $\cN=2$ CFT \cite{Henningson:1998gx} at leading order in $N$. In order for the black hole to have a smooth horizon, the charges must satisfy the non-linear constraint
\bea
 \label{chargeconstr}
 0=2 Q_2(Q_1+Q_2)^2 & + 2 a (Q_2^2+ 2 Q_1 Q_2) \\ -2 a(J_1+J_2)(Q_1+2 Q_2) 
 &- 2 a J_1 J_2 - 4 a^2(J_1+J_2)  \, .
\eea
The entropy \eqref{EntropyKNAdS5} can be written as the constrained Legendre transform of  the quantity \cite{Hosseini:2017mds}
\be\label{onshell}  \log {\cal Z} (X_i, \omega_i) =- 4 \pi \ii a \frac{X_1 X_2^2}{\omega_1\omega_2} \, .\ee
The entropy is obtained indeed by extremizing
\bea
S= \log {\cal Z} (X_i, \omega_i)  - 2 \pi \ii (\omega_1 J_1+\omega_2 J_2+ X_1 Q_1 +2 X_2 Q_2)\, , \nonumber
\eea
with respect to $X_i$ and $\omega_i$ with the constraint $X_1+2 X_2- \omega_1-\omega_2 =\pm 1$.
Both signs lead to the same critical value \eqref{EntropyKNAdS5}  \cite{Cabo-Bizet:2018ehj}, 
which is real precisely when \eqref{chargeconstr} is satisfied. 
The solution can be embedded in all AdS$_5$ type II or M-theory backgrounds preserving sixteen supercharges.

Examples of $\cN=2$ quiver theories  with a holographic dual and central charges of order $\cO(N^2)$ are provided by orbifolds of AdS$_5\times S^5$.  
Another large class of $\cN=2$ theories  with a holographic dual can be obtained by compactifying $N$ M5-branes on a Riemann surface $\Sigma_\fg$ of genus $\fg$ with regular punctures \cite{Gaiotto:2009gz}.
We will refer to these theories as holographic class $\cS$ theories.
The theories are generically non-Lagrangian and their central charge is of order $\cO(N^3)$.
On general grounds \cite{Gauntlett:2007sm,Cassani:2019vcl}, one knows that the effective 5d dual gravitational theory can be consistently truncated
to the minimal $\cN=4$ gauged supergravity.  As an example, one can check  explicitly that the universal KN solution can be embedded in the compactification with no punctures, corresponding to the M-theory AdS$_5$ solution originally found in \cite{Maldacena:2000mw}. To this purpose we can use the 5d consistent truncation of 7d $\U(1)^2$ gauged supergravity on $\Sigma_\fg$ derived in \cite{Szepietowski:2012tb,Cassani:2020cod,Malek:2020jsa}. The corresponding 5d theory   is written as an $\cN=2$ gauged supergravity with two vector multiplets and one hypermultiplet. At the AdS$_5$ vacuum one vector multiplet becomes massive through a Higgs  mechanism. The theory depends on a parameter $z$ that specifies the twist along $\Sigma_\fg$ and the preserved supersymmetry is generically $\cN=2$ corresponding to a dual $\cN=1$ CFT. For the special values $z=\pm 1$ supersymmetry is enhanced to $\cN=4$, and, for $\fg>1$, one obtains a dual $\cN=2$ CFT.  One can check that, for $z=\pm 1$ and $\fg>1$, by setting the hyperscalars to their AdS$_5$ value and setting to zero the fields in the massive vector multiplet, equations of motion and supersymmetry variations of the theory in \cite{Szepietowski:2012tb} coincide with those of the $\U(1)^2$ sector of  the minimal $\cN=4$ gauged supergravity. Explicitly, in the notations of \cite{Szepietowski:2012tb} we need to set  $e^{-10 B/3} = 2^{-1/3}  m^{10/3} \zeta$,  $ e^{10 \lambda_1/9}=e^{-5 \lambda_2/3}=2^{-1/3} \zeta$ and $A^{(0)}=m^{-4/3} 2^{-1/6} A$, with $g_2=2^{5/6}m^{5/3}$.

We will now show that the entropy \eqref{EntropyKNAdS5} matches the prediction of a microscopic computation based on the superconformal index for a generic $\cN=2$ CFT with a holographic dual.
The number of BPS states with charges $Q_1,Q_2$ and spin $J_1,J_2$  in an $\cN=2$ CFT can be computed (for large charges and spins) by taking the Legendre transform of the superconformal index $\cI$, which is a function
of chemical potentials for the electric charges and the angular momenta. The agreement between the gravitational picture and the field theory computation requires  $\log \cI = \log \cZ$.  The general expectation
for an $\cN=1$ CFT with a holographic dual is  \cite{Hosseini:2018dob}
\be\label{exp}  \log \cI(\Delta,\omega_i) =
- \frac{4 \pi \ii}{27} \frac{(\omega_1+\omega_2 \pm 1)^3}{\omega_1 \omega_2} a(\Delta) \, ,\ee
valid at large $N$. In this formula $\omega_1$ and $\omega_2$ are chemical potentials conjugated to $J_1$ and $J_2$ and $(\omega_1+\omega_2\pm 1) \Delta/2$ is a set of chemical potentials for the R- and flavor symmetries of the theory. 
They are normalized such that  $\Delta$ can be interpreted as an assignment of R-charges to the fields of the theory with the only constraint that the superpotential has R-charge two.
Moreover, $a(\Delta)=\frac{9}{32} \Tr R(\Delta)^3$, where $R(\Delta)$ is the R-symmetry generator and the trace is taken over the fermionic fields, is the trial $a$-charge at large $N$ \cite{Intriligator:2003jj}.
For Lagrangian theories  with a holographic dual, formula \eqref{exp} has been derived 
in the large $N$ limit in \cite{Benini:2018ywd,Lanir:2019abx,Benini:2020gjh,Cabo-Bizet:2020nkr}.
It is also compatible with the Cardy limit performed in \cite{Choi:2018hmj,Kim:2019yrz,Cabo-Bizet:2019osg,Amariti:2019mgp}.
It has not been yet derived in full generality for non-Lagrangian theories.
However, the Cardy limit can be also derived by writing the effective theory of the CFT
coupled to background fields on $S^3\times S^1$ in the limit where the circle shrinks \cite{Choi:2018hmj,Kim:2019yrz}
and this method applies also to non-Lagrangian theories.
The two signs in \eqref{exp} arise in the saddle point evaluation of the index in different regions in the space of chemical potentials.
Consider first an $\cN=2$  Lagrangian CFT with $n_{\text{V}}$ vector multiplets and $n_{\text{H}}$ hypermultiplets.
In $\cN=1$ language, the theory can be described by $n_{\text{V}}$ vector multiplets, $n_{\text{V}}$ chiral multiplets $\phi_I$
and $n_{\text{H}}$ pairs of chiral multiplets $(q_a,\tilde q_a)$.
The trial R-symmetry can be written as $R(\Delta)= (\Delta_1 r_1+\Delta_2 r_2)/2$ with $\Delta_1+\Delta_2=2$,
where  $r_1$ is the $\U(1)_R$ symmetry assigning charge $2$ to $\phi_I$ and zero to $q_a,\tilde q_a$
and  $r_2$  is the Cartan generator of $\SU(2)_R$ assigning charge zero to $\phi_I$ and charge $1$ to $q_a,\tilde q_a$.
In the gravitational dual,  $Q_1$ is the charge under $r_1/2$ and $Q_2$ under $r_2$.
Notice that the exact R-symmetry corresponds to $\hat\Delta_1=\frac23$, $\hat\Delta_2=\frac{4}{3}$ leading to canonical dimensions for the fields. We easily compute 
\bea
 \label{traces}
 \Tr R(\Delta) &= (n_{\text{V}} -n_{\text{H}})\Delta_1 \, , \\
  \Tr R(\Delta)^3 &= \frac{3 n_{\text{V}}}{4}  \Delta_1 \Delta_2^2 +\frac{n_{\text{V}} -n_{\text{H}}}{4} \Delta_1^3 \, .
\eea
Holography requires $a=c$  at large $N$ \cite{Henningson:1998gx}.
Using  $16(a-c) = \Tr R$ \cite{Anselmi:1997am}, this implies $n_{\text{V}}=n_{\text{H}}$ at leading order in $N$.
We then find  
\be\label{exp2}  \log \cI(\Delta,\omega_i) =
- \frac{ \pi \ii a}{8} \frac{(\omega_1+\omega_2 \pm1)^3}{\omega_1 \omega_2} \Delta_1 \Delta_2^2 \, ,\ee
where $a\equiv a(\hat\Delta)= \frac {n_{\text{V}}}{4}$ denotes the {\it exact} central charge of the CFT at large $N$. We see that,
with the redefinitions $X_1=(\omega_1+\omega_2\pm1)\frac{\Delta_1}{2}$ and $X_2=(\omega_1+\omega_2\pm1)\frac{\Delta_2}{4}$,  we recover \eqref{onshell} and the constraint
$X_1+2 X_2- \omega_1-\omega_2 = \pm 1$.  For non-Lagrangian theories, we  just  replace  $n_{\text{V}}$ and $n_{\text{H}}$ with an {\it effective} numbers of
vector multiplets and hypermultiplets, $n_\text{V}=4 (2 a - c), n_{\text{H}}=4 (5 c - 4 a)$ \cite{Gaiotto:2009gz}. The structure of \eqref{traces} is completely fixed by $\cN=2$ superconformal invariance \cite{Shapere:2008zf} and  the previous argument still holds. 

{\it The universal  black string in AdS$_5$}.--- We can similarly find a universal black string solution of $\cN=4$ gauged supergravity 
as a special case of  the black strings in AdS$_5\times S^5$ found in \cite{Hosseini:2019lkt} with angular momentum $J$, electric charges $(Q_1,Q_2,Q_3)$, and magnetic charges $(p_1,p_2,p_3)$ 
when we set $Q_3=Q_2$ and $p_3=p_2$. The horizon geometry of this solution has the topology of a warped product BTZ$\times_w S^2$ and carries an extra conserved charge corresponding
to a momentum $Q_0$ along the BTZ circle. Upon compactification on the circle we obtain a 4d dyonic black hole with Lifshitz-like asymptotics.
In the special case where all the electric charges $Q_i$ and $J$ are 
zero, the 5d solution is a domain wall interpolating between AdS$_5$ and AdS$_3\times S^2$.
There is a constraint on the magnetic charges $p_1 + 2 p_2 = - 1$,
which corresponds to the fact that the dual field theory is topologically twisted along $S^2$,
and a further constraint involving the electric charges: $( 1 + p_1 ) Q_2 + ( 1 + 2 p_1 ) Q_1 = 0$.
 The entropy of the four-dimensional  black hole reads \cite{Hosseini:2019lkt,Hosseini:2020vgl}
\be\label{entropyBS}
 S (J , Q_I, p_i) = 2 \pi \sqrt{ \frac{c_\text{CFT}}{6} \left(  Q_0 - \frac{J^2}{2 k} - \frac{F^2}{2 k_{FF}} \right)} \, ,
\ee
where $F=Q_1-Q_2$ and 
\bea\label{levels}
 & c_{\text{CFT}} = - 24 a \? \frac{( 1 + p_1 )^2}{2 + 3 p_1} \, , \\
 & k = - 2 a p_1 ( 1 + p_1 )^2 \, , \qquad k_{FF} = - 2 a ( 2 + 3 p_1 ) \, .
\eea
We expect again that the solution can be embedded in any AdS$_5$ compactification preserving sixteen supercharges. The special case of class $\cS$ with no punctures has been constructed 
in \cite{Hosseini:2020vgl}, as a part of a more general compactifications of M5-branes on a Riemann surface with eight supercharges. It is not difficult to check that the solution in \cite{Hosseini:2020vgl} becomes that of the universal black string when the  theory has sixteen supercharges ($\kappa =-1$, $z_1=-1\, , \fs_1=2-2 \fg$ and $\ft_1= -2 p_1$, in the notation of \cite{Hosseini:2020vgl}), and the entropy coincides with  \eqref{entropyBS}.

The result is indeed universal from the field theory point of view. The black string is dual to a 2d CFT obtained by compactifying the $\cN=2$ 4d CFT on  $S^2$ with a topological twist. The microscopic entropy is just the number of states
of the 2d CFT with $L_0=Q_0$, electric charges $Q_1$ and $Q_2$  under $r_1/2$ and $r_2$ and charge $J$ under the additional infrared $\SU(2)$ symmetry associated with rotation along $S^2$.
These states are accounted by  the charged Cardy formula. The latter has precisely the form  \eqref{entropyBS}, where $c_{\text{CFT}}$ is the central charge of the 2d CFT, $k$ is the level of the rotational $\SU(2)$ symmetry and $k_{FF}$ the level of the flavor symmetry $r_1/2-r_2$ \cite{Hosseini:2020vgl}. All these quantities can be computed with an (equivariant) integration of the 4d anomaly polynomial \cite{Benini:2013cda,Hosseini:2020vgl,Bah:2019rgq} and are universal because, as already noticed, the form of the four-dimensional anomalies for an $\cN=2$ CFT with a holographic dual is universal and only depends on $a=\frac{n_{\text{V}}}{4}$. For details on the anomaly integration leading to \eqref{levels} see \cite{Hosseini:2020vgl}.

{\it $\cN=4$ gauged supergravity in four-dimensions}.---%
4d $\cN = 4$ $\SO(4)$ gauged supergravity can be obtained as the consistent truncation of 11d supergravity on $S^7$ \cite{Cvetic:1999au}.
The bosonic field content of this theory is the metric $g_{\mu \nu}$, two $\SU(2)$ gauge fields $(A_{\mu}^I, \wt A^I_\mu)$, $I= 1,2,3$, a dilaton $\phi$ and an axion $\chi$.
The fermionic components are four gravitini $\psi_{\mu i}$, $i=1,2,3,4$, and four spin-$1/2$ fermions $\chi_i$.
These fields form the $\cN= 4$ gauged supergravity multiplet $(g_{\mu \nu}, \psi_{\mu i}, A_\mu^I, \wt A_\mu^I, \chi_i, \phi, \chi)$.
The bosonic Lagrangian is given by
\bea
 \frac{1}{\sqrt{- g}} & \cL = R - \frac12 (\pd^\mu \phi) (\pd_\mu \phi) - \frac12 e^{2 \phi} (\pd^\mu \chi) (\pd_\mu \chi) \\
 & + 2 g^2 \left( 4 + 2 \cosh(\phi) + \chi^2 e^{\phi} \right). \\
 & - \frac12 e^{- \phi} F_{\mu \nu}^I F^{I \mu \nu}
 - \frac12 \frac{e^{\phi}}{1 + \chi^2 e^{2 \phi}} \wt F_{\mu \nu}^{I} \wt F^{\mu \nu I} \\
 & - \frac{\chi}{2\sqrt{- g}} \varepsilon_{\mu \nu \rho \sigma} \Big( F^{\mu \nu I} F^{\rho \sigma I} - \frac{e^{2 \phi}}{1 + \chi^2 e^{2 \phi}} \wt F^{\mu \nu I} \wt F^{\rho \sigma I} \Big) \, . \nonumber
\eea

{\it The universal  KN black hole in AdS$_4$}.--- The universal 4d KN solution can be obtained by  specializing the general
KN black holes in AdS$_4 \times S^7$ to have angular momentum $J$, electric charges $(Q_1, Q_2, Q_1, Q_2)$ and magnetic charges $(p_1, p_2, p_1, p_2)$ under the Cartan subgroup of $\SO(8)$ \cite{Kostelecky:1995ei,Cvetic:2005zi,Hristov:2019mqp}.
The magnetic charges  are  restricted to satisfy $p_1 = - p_2 \equiv p$, which corresponds to the absence of a topological twist, and there is a further constraint among charges
\be
 \label{KN4-chargeconstr}
 J = \frac{Q_1 + Q_2}{2} \left( - 1 + \sqrt{1 - 16 p^2 + \frac{4 \pi^2} {F_ {S^3}^2} Q_{1} Q_{2}} \right ) \, .
\ee
The entropy, given by 
$S ( p , Q_1 , Q_2 ,J ) =  2 F_{S^3} \? \frac{J}{( Q_1 + Q_2 )}$,
where $F_{S^3} = \frac{\pi \ell_4^2}{2 G_{\text{N}}^{(4)}}$ is the $S^3$ free energy of the dual  CFT, 
can be obtained by extremizing the functional \cite{Choi:2018fdc,Hosseini:2019iad} 
\be
 S = - F_{S^3} \frac{ ( \Delta_1 \Delta_2 - 4 p^2 \omega^2 )}{\omega} + \pi \ii \sum_{i = 1}^2 \Delta_i Q_i +  \pi \ii \omega J  \, ,
\ee
with the constraint $\Delta_1 + \Delta_2 - \omega = 2$. This solution can be embedded in any AdS$_4$ solution of M-theory or type II string theory with sixteen supercharges.

The general expectation for a 3d  $\cN=4$ CFT with a holographic dual is that the entropy is the Legendre transform of \cite{Hosseini:2019iad}
\be
 \label{exp4d}
 \log \cI(\Delta,\omega) = - \frac{  F_{S^3}(\Delta_i- 2\omega p_i)}{2\omega} - \frac{  F_{S^3}(\Delta_i+ 2 \omega p_i)}{2 \omega} \, ,
\ee
where $F_{S^3}(\Delta_i)$ is the $S^3$ free energy as a function of the trial R-symmetry \cite{Jafferis:2010un} and $\Delta_1+\Delta_2=2$. $\Delta_1$ and $\Delta_2$ are conjugated to the Cartan generators of the $\SU(2)\times \SU(2)$  R-symmetry.
This formula follows from gluing gravitational blocks in gravity \cite{Hosseini:2019iad}, which is the counterpart of gluing holomorphic blocks in the dual field theory \cite{Pasquetti:2011fj,Beem:2012mb}. For $p=0$ and the ABJM theory, \eqref{exp4d} has been derived in the Cardy limit
in \cite{Choi:2019zpz,Choi:2019dfu} by factorizing the superconformal index into vortex partition functions. It is expected to hold for more general theories and for $p\ne 0$.  
We then find a general {\it prediction} for the trial free energy of a generic $\cN=4$ CFT with a holographic dual in the large $N$ limit
\be \label{prediction} F_{S^3}(\Delta_i) = F_{S^3} \Delta_1 \Delta_2 \, .\ee

We can explicitly check this prediction in various examples.
Holographic $\cN=4$ CFTs arise as woldvolume theories of M2-branes probing $\mathbb{C}^2/\Gamma_1\times \mathbb{C}^2/\Gamma_2$, with $\Gamma_i$ discrete subgroups of $\SU(2)$, where the role of $\Gamma_1$ and $\Gamma_2$ can be exchanged by mirror symmetry \cite{Porrati:1996xi}. Consider, for simplicity, the case where $\Gamma_2=\mathbb{Z}_{p}$.
The worldvolume theory is based on an $\cN=4$ ADE quiver  with gauge groups $\U(n_a N)$ corresponding to the nodes of the extended Dynkin diagram of $\Gamma_1$     and bifundamental hypermultiplets associated to the links, flavored with the addition of $p$ fundamental hypermultiplets  ($n_a$ are the co-marks: see \cite{Mekareeya:2015bla} for conventions and details).    Denote by $n_{\text{V}}$, $n_{\text{B}}$ and $n_{\text{F}}$  the total number of vector multiplets,  bi-fundamental   and  fundamental hypers, respectively.  The large $N$ limit of the $S^3$ partition function can be computed with the methods in \cite{Herzog:2010hf,Jafferis:2011zi,Crichigno:2012sk}.  In the large $N$ limit the eigenvalue distribution for a group $\U(n_a N)$  is given by  $n_a$ copies of the segment  $\lambda(t)=N^{1/2} t$ with density $\rho(t)$ ($\int dt \rho(t)=1$). 
In the large $N$ limit, using the rules in  \cite{Jafferis:2011zi,Crichigno:2012sk},  we obtain
\bea
 F_{S^3}(\Delta_i) &= \frac{n_{\text{B}}}{N^{1/2}}  \frac{\pi^2}{6}   \Delta_2(\Delta_2-2)(\Delta_2-4)  \int \rho(t)^2 dt \\&
 +  \frac{n_{\text{V}}}{N^{1/2}}  \frac{2\pi^2}{3}  \Delta_1(\Delta_1-1)(\Delta_1-2) \int \rho(t)^2 dt \\&
 +   \frac{n_{\text{F}} N^{1/2}}{2} (2-\Delta_2) \int \rho(t) |t| dt  \, ,
\eea 
where we assign charge $\Delta_1$  to the adjoint chiral in the vector multiplet 
and  $\Delta_2/2$ to the chiral fields $q_a,\tilde q_a$ in the hypermultiplets, in $\cN=2$ notations.
Since   the ADE quivers are balanced (the number of  hypers for each group is twice the number of colors), we  have $n_{\text{V}}=n_{\text{B}}$  and we find the saddle-point distribution
\be  \rho(t) =\frac{\pi \sqrt{2 n_{\text{F}} n_{\text{V}} N} \Delta_2 - n_{\text{F}} N |t|}{2 \pi^2 n_{\text{V}} \Delta_2^2}\, ,\ee
with free energy 
\be\label{M2} F_{S^3}(\Delta_i) =  \frac{\pi}{3} \sqrt{2 n_{\text{F}} n_{\text{V}}} \Delta_1\Delta_2 \, ,\ee
which reproduces \eqref{prediction} with $F_{S^3}= \frac{\pi}{3}\sqrt{2 n_{\text{F}} n_{\text{V}}}$. The previous computation for $\Gamma_1=\mathbb{Z}_q$ was already done in disguise in \cite{Hosseini:2016ume}. Notice that $F_{S^3}=\cO(N^{3/2})$, as expected for M2-brane theories. Formula \eqref{M2} can be also derived from the identification of the trial free energy with the volume functional of the transverse Calabi-Yau \cite{Jafferis:2011zi,Martelli:2006yb}.  M2 branes probing abelian hyperk\"ahler orbifolds of $\mathbb{C}^4$ can be also realized  in terms of  $\cN=4$  circular quivers with non-zero Chern-Simons terms \cite{Imamura:2008nn}.  The simplest example is actually  ABJM itself, whose free energy  has been computed in \cite{Jafferis:2011zi} and reads $F_{S^3}(\delta_i) = 4 F_{S^3} \sqrt{\delta_1\delta_2\delta_3\delta_4}$, where $\delta_i$ are conjugated to the Cartan subgroup of $\SO(8)$ and satisfy $\sum_{i=1}^4 \delta_i=2$. This reduces to \eqref{prediction} for $\delta_3=\delta_1=\Delta_1/2$ and  $\delta_4=\delta_2=\Delta_2/2$. Using and extending the results in \cite{Hosseini:2016ume,Amariti:2011uw}, one can also compute the free energy for the more general $\cN=4$ quivers discussed in \cite{Imamura:2008nn} and check that \eqref{prediction} is valid.
Notice that in all these examples  the $\SU(2)\times \SU(2)$  R-symmetry  acts differently from the  case with no Chern-Simons, and is fully visible once the theory is written in terms of both standard and twisted hypermultiplets \cite{Gaiotto:2008sd,Imamura:2008dt,Hosomichi:2008jd}. Another large class of $\cN=4$ holographic quivers are the $T^\rho_\sigma(G)$ theories \cite{Gaiotto:2008sa} whose gravitational dual was found in \cite{Assel:2011xz,Assel:2012cj}.
The refined free energy on $S^3$ for $T(\SU(N))$ has been recently computed in the large $N$ limit  in \cite{Coccia:2020cku} and it reads $F_{S^3}= \frac12 \Delta_1\Delta_2 N^2 \log N$ which also agrees with \eqref{prediction}. 
The prediction \eqref{prediction} can be also checked for a larger class of $T^\rho_\sigma(G)$ theories \cite{Coccia:2020wtk}.

{\it The universal  twisted  black hole in AdS$_4$}.--- We can obtain dyonic  black holes with a twist (magnetic charge for the R-symmetry) and horizon AdS$_2\times \Sigma_\fg$, where $\Sigma_\fg$ is a Riemann surface of genus $\fg$, by specializing the corresponding solution in AdS$_4 \times S^7$ \cite{Cacciatori:2009iz,Katmadas:2014faa,Halmagyi:2014qza}. For $\fg=0$ we can add an angular momentum $J$ \cite{Hristov:2018spe}. The case of static solutions of minimal gauged supergravity has been already discussed in \cite{Azzurli:2017kxo,Bobev:2017uzs}. Solutions with a generic  $\cN=4$ choice of charges are not regular  and we will not discuss them further. One can check however that the comparison between the gravity entropy functional and the large $N$ limit of the (refined) topologically twisted index  would still agree (since it is also based on \eqref{prediction} \cite{Hosseini:2019iad}), although the extremization leads to a non-physical value for the entropy.
For $J=0$ one can find an off-shell agreement by considering the Euclidean black saddles discussed in \cite{Bobev:2020pjk}.

\begin{acknowledgements}
SMH is supported in part by WPI Initiative, MEXT, Japan at IPMU, the University of Tokyo, JSPS KAKENHI Grant-in-Aid (Wakate-A), No.17H04837 and JSPS KAKENHI Grant-in-Aid (Early-Career Scientists), No.20K14462.
AZ is partially supported by the INFN, the ERC-STG grant 637844-HBQFTNCER, and the MIUR-PRIN contract 2017CC72MK003.
\end{acknowledgements}

\bibliography{UnivBHs}

\begin{thebibliography}{79}%
\makeatletter
\providecommand \@ifxundefined [1]{%
 \@ifx{#1\undefined}
}%
\providecommand \@ifnum [1]{%
 \ifnum #1\expandafter \@firstoftwo
 \else \expandafter \@secondoftwo
 \fi
}%
\providecommand \@ifx [1]{%
 \ifx #1\expandafter \@firstoftwo
 \else \expandafter \@secondoftwo
 \fi
}%
\providecommand \natexlab [1]{#1}%
\providecommand \enquote  [1]{``#1''}%
\providecommand \bibnamefont  [1]{#1}%
\providecommand \bibfnamefont [1]{#1}%
\providecommand \citenamefont [1]{#1}%
\providecommand \href@noop [0]{\@secondoftwo}%
\providecommand \href [0]{\begingroup \@sanitize@url \@href}%
\providecommand \@href[1]{\@@startlink{#1}\@@href}%
\providecommand \@@href[1]{\endgroup#1\@@endlink}%
\providecommand \@sanitize@url [0]{\catcode `\\12\catcode `\$12\catcode
  `\&12\catcode `\#12\catcode `\^12\catcode `\_12\catcode `\%12\relax}%
\providecommand \@@startlink[1]{}%
\providecommand \@@endlink[0]{}%
\providecommand \url  [0]{\begingroup\@sanitize@url \@url }%
\providecommand \@url [1]{\endgroup\@href {#1}{\urlprefix }}%
\providecommand \urlprefix  [0]{URL }%
\providecommand \Eprint [0]{\href }%
\providecommand \doibase [0]{http://dx.doi.org/}%
\providecommand \selectlanguage [0]{\@gobble}%
\providecommand \bibinfo  [0]{\@secondoftwo}%
\providecommand \bibfield  [0]{\@secondoftwo}%
\providecommand \translation [1]{[#1]}%
\providecommand \BibitemOpen [0]{}%
\providecommand \bibitemStop [0]{}%
\providecommand \bibitemNoStop [0]{.\EOS\space}%
\providecommand \EOS [0]{\spacefactor3000\relax}%
\providecommand \BibitemShut  [1]{\csname bibitem#1\endcsname}%
\let\auto@bib@innerbib\@empty
\bibitem [{\citenamefont {Strominger}\ and\ \citenamefont
  {Vafa}(1996)}]{Strominger:1996sh}%
  \BibitemOpen
  \bibfield  {author} {\bibinfo {author} {\bibfnamefont {A.}~\bibnamefont
  {Strominger}}\ and\ \bibinfo {author} {\bibfnamefont {C.}~\bibnamefont
  {Vafa}},\ }\href {\doibase 10.1016/0370-2693(96)00345-0} {\bibfield
  {journal} {\bibinfo  {journal} {Phys. Lett.}\ }\textbf {\bibinfo {volume}
  {B379}},\ \bibinfo {pages} {99} (\bibinfo {year} {1996})},\ \Eprint
  {http://arxiv.org/abs/hep-th/9601029} {arXiv:hep-th/9601029 [hep-th]}
  \BibitemShut {NoStop}%
\bibitem [{\citenamefont {Maldacena}(1999)}]{Maldacena:1997re}%
  \BibitemOpen
  \bibfield  {author} {\bibinfo {author} {\bibfnamefont {J.~M.}\ \bibnamefont
  {Maldacena}},\ }\href {\doibase 10.1023/A:1026654312961} {\bibfield
  {journal} {\bibinfo  {journal} {Int. J. Theor. Phys.}\ }\textbf {\bibinfo
  {volume} {38}},\ \bibinfo {pages} {1113} (\bibinfo {year} {1999})},\ \Eprint
  {http://arxiv.org/abs/hep-th/9711200} {arXiv:hep-th/9711200} \BibitemShut
  {NoStop}%
\bibitem [{\citenamefont {Benini}\ \emph
  {et~al.}(2016{\natexlab{a}})\citenamefont {Benini}, \citenamefont {Hristov},\
  and\ \citenamefont {Zaffaroni}}]{Benini:2015eyy}%
  \BibitemOpen
  \bibfield  {author} {\bibinfo {author} {\bibfnamefont {F.}~\bibnamefont
  {Benini}}, \bibinfo {author} {\bibfnamefont {K.}~\bibnamefont {Hristov}}, \
  and\ \bibinfo {author} {\bibfnamefont {A.}~\bibnamefont {Zaffaroni}},\ }\href
  {\doibase 10.1007/JHEP05(2016)054} {\bibfield  {journal} {\bibinfo  {journal}
  {JHEP}\ }\textbf {\bibinfo {volume} {05}},\ \bibinfo {pages} {054} (\bibinfo
  {year} {2016}{\natexlab{a}})},\ \Eprint {http://arxiv.org/abs/1511.04085}
  {arXiv:1511.04085 [hep-th]} \BibitemShut {NoStop}%
\bibitem [{\citenamefont {Benini}\ and\ \citenamefont
  {Milan}(2020)}]{Benini:2018ywd}%
  \BibitemOpen
  \bibfield  {author} {\bibinfo {author} {\bibfnamefont {F.}~\bibnamefont
  {Benini}}\ and\ \bibinfo {author} {\bibfnamefont {P.}~\bibnamefont {Milan}},\
  }\href {\doibase 10.1103/PhysRevX.10.021037} {\bibfield  {journal} {\bibinfo
  {journal} {Phys. Rev. X}\ }\textbf {\bibinfo {volume} {10}},\ \bibinfo
  {pages} {021037} (\bibinfo {year} {2020})},\ \Eprint
  {http://arxiv.org/abs/1812.09613} {arXiv:1812.09613 [hep-th]} \BibitemShut
  {NoStop}%
\bibitem [{\citenamefont {Choi}\ \emph {et~al.}(2018)\citenamefont {Choi},
  \citenamefont {Kim}, \citenamefont {Kim},\ and\ \citenamefont
  {Nahmgoong}}]{Choi:2018hmj}%
  \BibitemOpen
  \bibfield  {author} {\bibinfo {author} {\bibfnamefont {S.}~\bibnamefont
  {Choi}}, \bibinfo {author} {\bibfnamefont {J.}~\bibnamefont {Kim}}, \bibinfo
  {author} {\bibfnamefont {S.}~\bibnamefont {Kim}}, \ and\ \bibinfo {author}
  {\bibfnamefont {J.}~\bibnamefont {Nahmgoong}},\ }\href@noop {} {\  (\bibinfo
  {year} {2018})},\ \Eprint {http://arxiv.org/abs/1810.12067} {arXiv:1810.12067
  [hep-th]} \BibitemShut {NoStop}%
\bibitem [{\citenamefont {Pestun}\ \emph {et~al.}(2017)\citenamefont {Pestun}
  \emph {et~al.}}]{Pestun:2016zxk}%
  \BibitemOpen
  \bibfield  {author} {\bibinfo {author} {\bibfnamefont {V.}~\bibnamefont
  {Pestun}} \emph {et~al.},\ }\href {\doibase 10.1088/1751-8121/aa63c1}
  {\bibfield  {journal} {\bibinfo  {journal} {J. Phys. A}\ }\textbf {\bibinfo
  {volume} {50}},\ \bibinfo {pages} {440301} (\bibinfo {year} {2017})},\
  \Eprint {http://arxiv.org/abs/1608.02952} {arXiv:1608.02952 [hep-th]}
  \BibitemShut {NoStop}%
\bibitem [{\citenamefont {Azzurli}\ \emph {et~al.}(2018)\citenamefont
  {Azzurli}, \citenamefont {Bobev}, \citenamefont {Crichigno}, \citenamefont
  {Min},\ and\ \citenamefont {Zaffaroni}}]{Azzurli:2017kxo}%
  \BibitemOpen
  \bibfield  {author} {\bibinfo {author} {\bibfnamefont {F.}~\bibnamefont
  {Azzurli}}, \bibinfo {author} {\bibfnamefont {N.}~\bibnamefont {Bobev}},
  \bibinfo {author} {\bibfnamefont {P.~M.}\ \bibnamefont {Crichigno}}, \bibinfo
  {author} {\bibfnamefont {V.~S.}\ \bibnamefont {Min}}, \ and\ \bibinfo
  {author} {\bibfnamefont {A.}~\bibnamefont {Zaffaroni}},\ }\href {\doibase
  10.1007/JHEP02(2018)054} {\bibfield  {journal} {\bibinfo  {journal} {JHEP}\
  }\textbf {\bibinfo {volume} {02}},\ \bibinfo {pages} {054} (\bibinfo {year}
  {2018})},\ \Eprint {http://arxiv.org/abs/1707.04257} {arXiv:1707.04257
  [hep-th]} \BibitemShut {NoStop}%
\bibitem [{\citenamefont {Cabo-Bizet}\ \emph
  {et~al.}(2019{\natexlab{a}})\citenamefont {Cabo-Bizet}, \citenamefont
  {Cassani}, \citenamefont {Martelli},\ and\ \citenamefont
  {Murthy}}]{Cabo-Bizet:2019osg}%
  \BibitemOpen
  \bibfield  {author} {\bibinfo {author} {\bibfnamefont {A.}~\bibnamefont
  {Cabo-Bizet}}, \bibinfo {author} {\bibfnamefont {D.}~\bibnamefont {Cassani}},
  \bibinfo {author} {\bibfnamefont {D.}~\bibnamefont {Martelli}}, \ and\
  \bibinfo {author} {\bibfnamefont {S.}~\bibnamefont {Murthy}},\ }\href
  {\doibase 10.1007/JHEP08(2019)120} {\bibfield  {journal} {\bibinfo  {journal}
  {JHEP}\ }\textbf {\bibinfo {volume} {08}},\ \bibinfo {pages} {120} (\bibinfo
  {year} {2019}{\natexlab{a}})},\ \Eprint {http://arxiv.org/abs/1904.05865}
  {arXiv:1904.05865 [hep-th]} \BibitemShut {NoStop}%
\bibitem [{\citenamefont {Kim}\ \emph {et~al.}(2019)\citenamefont {Kim},
  \citenamefont {Kim},\ and\ \citenamefont {Song}}]{Kim:2019yrz}%
  \BibitemOpen
  \bibfield  {author} {\bibinfo {author} {\bibfnamefont {J.}~\bibnamefont
  {Kim}}, \bibinfo {author} {\bibfnamefont {S.}~\bibnamefont {Kim}}, \ and\
  \bibinfo {author} {\bibfnamefont {J.}~\bibnamefont {Song}},\ }\href@noop {}
  {\  (\bibinfo {year} {2019})},\ \Eprint {http://arxiv.org/abs/1904.03455}
  {arXiv:1904.03455 [hep-th]} \BibitemShut {NoStop}%
\bibitem [{\citenamefont {Benini}\ \emph {et~al.}(2020)\citenamefont {Benini},
  \citenamefont {Colombo}, \citenamefont {Soltani}, \citenamefont {Zaffaroni},\
  and\ \citenamefont {Zhang}}]{Benini:2020gjh}%
  \BibitemOpen
  \bibfield  {author} {\bibinfo {author} {\bibfnamefont {F.}~\bibnamefont
  {Benini}}, \bibinfo {author} {\bibfnamefont {E.}~\bibnamefont {Colombo}},
  \bibinfo {author} {\bibfnamefont {S.}~\bibnamefont {Soltani}}, \bibinfo
  {author} {\bibfnamefont {A.}~\bibnamefont {Zaffaroni}}, \ and\ \bibinfo
  {author} {\bibfnamefont {Z.}~\bibnamefont {Zhang}},\ }\href@noop {} {\
  (\bibinfo {year} {2020})},\ \Eprint {http://arxiv.org/abs/2005.12308}
  {arXiv:2005.12308 [hep-th]} \BibitemShut {NoStop}%
\bibitem [{\citenamefont {Couzens}\ \emph {et~al.}(2019)\citenamefont
  {Couzens}, \citenamefont {Gauntlett}, \citenamefont {Martelli},\ and\
  \citenamefont {Sparks}}]{Couzens:2018wnk}%
  \BibitemOpen
  \bibfield  {author} {\bibinfo {author} {\bibfnamefont {C.}~\bibnamefont
  {Couzens}}, \bibinfo {author} {\bibfnamefont {J.~P.}\ \bibnamefont
  {Gauntlett}}, \bibinfo {author} {\bibfnamefont {D.}~\bibnamefont {Martelli}},
  \ and\ \bibinfo {author} {\bibfnamefont {J.}~\bibnamefont {Sparks}},\ }\href
  {\doibase 10.1007/JHEP01(2019)212} {\bibfield  {journal} {\bibinfo  {journal}
  {JHEP}\ }\textbf {\bibinfo {volume} {01}},\ \bibinfo {pages} {212} (\bibinfo
  {year} {2019})},\ \Eprint {http://arxiv.org/abs/1810.11026} {arXiv:1810.11026
  [hep-th]} \BibitemShut {NoStop}%
\bibitem [{\citenamefont {Gauntlett}\ \emph {et~al.}(2019)\citenamefont
  {Gauntlett}, \citenamefont {Martelli},\ and\ \citenamefont
  {Sparks}}]{Gauntlett:2019pqg}%
  \BibitemOpen
  \bibfield  {author} {\bibinfo {author} {\bibfnamefont {J.~P.}\ \bibnamefont
  {Gauntlett}}, \bibinfo {author} {\bibfnamefont {D.}~\bibnamefont {Martelli}},
  \ and\ \bibinfo {author} {\bibfnamefont {J.}~\bibnamefont {Sparks}},\ }\href
  {\doibase 10.1007/JHEP11(2019)176} {\bibfield  {journal} {\bibinfo  {journal}
  {JHEP}\ }\textbf {\bibinfo {volume} {11}},\ \bibinfo {pages} {176} (\bibinfo
  {year} {2019})},\ \Eprint {http://arxiv.org/abs/1910.08078} {arXiv:1910.08078
  [hep-th]} \BibitemShut {NoStop}%
\bibitem [{\citenamefont {Gauntlett}\ and\ \citenamefont
  {Varela}(2007)}]{Gauntlett:2007ma}%
  \BibitemOpen
  \bibfield  {author} {\bibinfo {author} {\bibfnamefont {J.~P.}\ \bibnamefont
  {Gauntlett}}\ and\ \bibinfo {author} {\bibfnamefont {O.}~\bibnamefont
  {Varela}},\ }\href {\doibase 10.1103/PhysRevD.76.126007} {\bibfield
  {journal} {\bibinfo  {journal} {Phys. Rev. D}\ }\textbf {\bibinfo {volume}
  {76}},\ \bibinfo {pages} {126007} (\bibinfo {year} {2007})},\ \Eprint
  {http://arxiv.org/abs/0707.2315} {arXiv:0707.2315 [hep-th]} \BibitemShut
  {NoStop}%
\bibitem [{\citenamefont {Gauntlett}\ and\ \citenamefont
  {Varela}(2008)}]{Gauntlett:2007sm}%
  \BibitemOpen
  \bibfield  {author} {\bibinfo {author} {\bibfnamefont {J.~P.}\ \bibnamefont
  {Gauntlett}}\ and\ \bibinfo {author} {\bibfnamefont {O.}~\bibnamefont
  {Varela}},\ }\href {\doibase 10.1088/1126-6708/2008/02/083} {\bibfield
  {journal} {\bibinfo  {journal} {JHEP}\ }\textbf {\bibinfo {volume} {02}},\
  \bibinfo {pages} {083} (\bibinfo {year} {2008})},\ \Eprint
  {http://arxiv.org/abs/0712.3560} {arXiv:0712.3560 [hep-th]} \BibitemShut
  {NoStop}%
\bibitem [{\citenamefont {Cassani}\ \emph {et~al.}(2019)\citenamefont
  {Cassani}, \citenamefont {Josse}, \citenamefont {Petrini},\ and\
  \citenamefont {Waldram}}]{Cassani:2019vcl}%
  \BibitemOpen
  \bibfield  {author} {\bibinfo {author} {\bibfnamefont {D.}~\bibnamefont
  {Cassani}}, \bibinfo {author} {\bibfnamefont {G.}~\bibnamefont {Josse}},
  \bibinfo {author} {\bibfnamefont {M.}~\bibnamefont {Petrini}}, \ and\
  \bibinfo {author} {\bibfnamefont {D.}~\bibnamefont {Waldram}},\ }\href
  {\doibase 10.1007/JHEP11(2019)017} {\bibfield  {journal} {\bibinfo  {journal}
  {JHEP}\ }\textbf {\bibinfo {volume} {11}},\ \bibinfo {pages} {017} (\bibinfo
  {year} {2019})},\ \Eprint {http://arxiv.org/abs/1907.06730} {arXiv:1907.06730
  [hep-th]} \BibitemShut {NoStop}%
\bibitem [{\citenamefont {Malek}(2017)}]{Malek:2017njj}%
  \BibitemOpen
  \bibfield  {author} {\bibinfo {author} {\bibfnamefont {E.}~\bibnamefont
  {Malek}},\ }\href {\doibase 10.1002/prop.201700061} {\bibfield  {journal}
  {\bibinfo  {journal} {Fortsch. Phys.}\ }\textbf {\bibinfo {volume} {65}},\
  \bibinfo {pages} {1700061} (\bibinfo {year} {2017})},\ \Eprint
  {http://arxiv.org/abs/1707.00714} {arXiv:1707.00714 [hep-th]} \BibitemShut
  {NoStop}%
\bibitem [{\citenamefont {Benini}\ \emph
  {et~al.}(2016{\natexlab{b}})\citenamefont {Benini}, \citenamefont {Bobev},\
  and\ \citenamefont {Crichigno}}]{Benini:2015bwz}%
  \BibitemOpen
  \bibfield  {author} {\bibinfo {author} {\bibfnamefont {F.}~\bibnamefont
  {Benini}}, \bibinfo {author} {\bibfnamefont {N.}~\bibnamefont {Bobev}}, \
  and\ \bibinfo {author} {\bibfnamefont {P.~M.}\ \bibnamefont {Crichigno}},\
  }\href {\doibase 10.1007/JHEP07(2016)020} {\bibfield  {journal} {\bibinfo
  {journal} {JHEP}\ }\textbf {\bibinfo {volume} {07}},\ \bibinfo {pages} {020}
  (\bibinfo {year} {2016}{\natexlab{b}})},\ \Eprint
  {http://arxiv.org/abs/1511.09462} {arXiv:1511.09462 [hep-th]} \BibitemShut
  {NoStop}%
\bibitem [{\citenamefont {Bobev}\ and\ \citenamefont
  {Crichigno}(2017)}]{Bobev:2017uzs}%
  \BibitemOpen
  \bibfield  {author} {\bibinfo {author} {\bibfnamefont {N.}~\bibnamefont
  {Bobev}}\ and\ \bibinfo {author} {\bibfnamefont {P.~M.}\ \bibnamefont
  {Crichigno}},\ }\href {\doibase 10.1007/JHEP12(2017)065} {\bibfield
  {journal} {\bibinfo  {journal} {JHEP}\ }\textbf {\bibinfo {volume} {12}},\
  \bibinfo {pages} {065} (\bibinfo {year} {2017})},\ \Eprint
  {http://arxiv.org/abs/1708.05052} {arXiv:1708.05052 [hep-th]} \BibitemShut
  {NoStop}%
\bibitem [{\citenamefont {Bobev}\ and\ \citenamefont
  {Crichigno}(2019)}]{Bobev:2019zmz}%
  \BibitemOpen
  \bibfield  {author} {\bibinfo {author} {\bibfnamefont {N.}~\bibnamefont
  {Bobev}}\ and\ \bibinfo {author} {\bibfnamefont {P.~M.}\ \bibnamefont
  {Crichigno}},\ }\href {\doibase 10.1007/JHEP12(2019)054} {\bibfield
  {journal} {\bibinfo  {journal} {JHEP}\ }\textbf {\bibinfo {volume} {12}},\
  \bibinfo {pages} {054} (\bibinfo {year} {2019})},\ \Eprint
  {http://arxiv.org/abs/1909.05873} {arXiv:1909.05873 [hep-th]} \BibitemShut
  {NoStop}%
\bibitem [{\citenamefont {Romans}(1986)}]{Romans:1985ps}%
  \BibitemOpen
  \bibfield  {author} {\bibinfo {author} {\bibfnamefont {L.}~\bibnamefont
  {Romans}},\ }\href {\doibase 10.1016/0550-3213(86)90398-6} {\bibfield
  {journal} {\bibinfo  {journal} {Nucl. Phys. B}\ }\textbf {\bibinfo {volume}
  {267}},\ \bibinfo {pages} {433} (\bibinfo {year} {1986})}\BibitemShut
  {NoStop}%
\bibitem [{\citenamefont {Awada}\ and\ \citenamefont
  {Townsend}(1985)}]{Awada:1985ep}%
  \BibitemOpen
  \bibfield  {author} {\bibinfo {author} {\bibfnamefont {M.}~\bibnamefont
  {Awada}}\ and\ \bibinfo {author} {\bibfnamefont {P.}~\bibnamefont
  {Townsend}},\ }\href {\doibase 10.1016/0550-3213(85)90156-7} {\bibfield
  {journal} {\bibinfo  {journal} {Nucl. Phys. B}\ }\textbf {\bibinfo {volume}
  {255}},\ \bibinfo {pages} {617} (\bibinfo {year} {1985})}\BibitemShut
  {NoStop}%
\bibitem [{\citenamefont {Gutowski}\ and\ \citenamefont
  {Reall}(2004{\natexlab{a}})}]{Gutowski:2004ez}%
  \BibitemOpen
  \bibfield  {author} {\bibinfo {author} {\bibfnamefont {J.~B.}\ \bibnamefont
  {Gutowski}}\ and\ \bibinfo {author} {\bibfnamefont {H.~S.}\ \bibnamefont
  {Reall}},\ }\href {\doibase 10.1088/1126-6708/2004/02/006} {\bibfield
  {journal} {\bibinfo  {journal} {JHEP}\ }\textbf {\bibinfo {volume} {02}},\
  \bibinfo {pages} {006} (\bibinfo {year} {2004}{\natexlab{a}})},\ \Eprint
  {http://arxiv.org/abs/hep-th/0401042} {arXiv:hep-th/0401042 [hep-th]}
  \BibitemShut {NoStop}%
\bibitem [{\citenamefont {Gutowski}\ and\ \citenamefont
  {Reall}(2004{\natexlab{b}})}]{Gutowski:2004yv}%
  \BibitemOpen
  \bibfield  {author} {\bibinfo {author} {\bibfnamefont {J.~B.}\ \bibnamefont
  {Gutowski}}\ and\ \bibinfo {author} {\bibfnamefont {H.~S.}\ \bibnamefont
  {Reall}},\ }\href {\doibase 10.1088/1126-6708/2004/04/048} {\bibfield
  {journal} {\bibinfo  {journal} {JHEP}\ }\textbf {\bibinfo {volume} {04}},\
  \bibinfo {pages} {048} (\bibinfo {year} {2004}{\natexlab{b}})},\ \Eprint
  {http://arxiv.org/abs/hep-th/0401129} {arXiv:hep-th/0401129 [hep-th]}
  \BibitemShut {NoStop}%
\bibitem [{\citenamefont {Chong}\ \emph
  {et~al.}(2005{\natexlab{a}})\citenamefont {Chong}, \citenamefont
  {Cveti\v{c}}, \citenamefont {Lu},\ and\ \citenamefont {Pope}}]{Chong:2005da}%
  \BibitemOpen
  \bibfield  {author} {\bibinfo {author} {\bibfnamefont {Z.~W.}\ \bibnamefont
  {Chong}}, \bibinfo {author} {\bibfnamefont {M.}~\bibnamefont {Cveti\v{c}}},
  \bibinfo {author} {\bibfnamefont {H.}~\bibnamefont {Lu}}, \ and\ \bibinfo
  {author} {\bibfnamefont {C.~N.}\ \bibnamefont {Pope}},\ }\href {\doibase
  10.1103/PhysRevD.72.041901} {\bibfield  {journal} {\bibinfo  {journal} {Phys.
  Rev. D}\ }\textbf {\bibinfo {volume} {72}},\ \bibinfo {pages} {041901}
  (\bibinfo {year} {2005}{\natexlab{a}})},\ \Eprint
  {http://arxiv.org/abs/hep-th/0505112} {arXiv:hep-th/0505112 [hep-th]}
  \BibitemShut {NoStop}%
\bibitem [{\citenamefont {Chong}\ \emph
  {et~al.}(2005{\natexlab{b}})\citenamefont {Chong}, \citenamefont
  {Cveti\v{c}}, \citenamefont {Lu},\ and\ \citenamefont {Pope}}]{Chong:2005hr}%
  \BibitemOpen
  \bibfield  {author} {\bibinfo {author} {\bibfnamefont {Z.-W.}\ \bibnamefont
  {Chong}}, \bibinfo {author} {\bibfnamefont {M.}~\bibnamefont {Cveti\v{c}}},
  \bibinfo {author} {\bibfnamefont {H.}~\bibnamefont {Lu}}, \ and\ \bibinfo
  {author} {\bibfnamefont {C.~N.}\ \bibnamefont {Pope}},\ }\href {\doibase
  10.1103/PhysRevLett.95.161301} {\bibfield  {journal} {\bibinfo  {journal}
  {Phys. Rev. Lett.}\ }\textbf {\bibinfo {volume} {95}},\ \bibinfo {pages}
  {161301} (\bibinfo {year} {2005}{\natexlab{b}})},\ \Eprint
  {http://arxiv.org/abs/hep-th/0506029} {arXiv:hep-th/0506029 [hep-th]}
  \BibitemShut {NoStop}%
\bibitem [{\citenamefont {Kunduri}\ \emph {et~al.}(2006)\citenamefont
  {Kunduri}, \citenamefont {Lucietti},\ and\ \citenamefont
  {Reall}}]{Kunduri:2006ek}%
  \BibitemOpen
  \bibfield  {author} {\bibinfo {author} {\bibfnamefont {H.~K.}\ \bibnamefont
  {Kunduri}}, \bibinfo {author} {\bibfnamefont {J.}~\bibnamefont {Lucietti}}, \
  and\ \bibinfo {author} {\bibfnamefont {H.~S.}\ \bibnamefont {Reall}},\ }\href
  {\doibase 10.1088/1126-6708/2006/04/036} {\bibfield  {journal} {\bibinfo
  {journal} {JHEP}\ }\textbf {\bibinfo {volume} {04}},\ \bibinfo {pages} {036}
  (\bibinfo {year} {2006})},\ \Eprint {http://arxiv.org/abs/hep-th/0601156}
  {arXiv:hep-th/0601156 [hep-th]} \BibitemShut {NoStop}%
\bibitem [{\citenamefont {Kim}\ and\ \citenamefont {Lee}(2006)}]{Kim:2006he}%
  \BibitemOpen
  \bibfield  {author} {\bibinfo {author} {\bibfnamefont {S.}~\bibnamefont
  {Kim}}\ and\ \bibinfo {author} {\bibfnamefont {K.-M.}\ \bibnamefont {Lee}},\
  }\href {\doibase 10.1088/1126-6708/2006/12/077} {\bibfield  {journal}
  {\bibinfo  {journal} {JHEP}\ }\textbf {\bibinfo {volume} {12}},\ \bibinfo
  {pages} {077} (\bibinfo {year} {2006})},\ \Eprint
  {http://arxiv.org/abs/hep-th/0607085} {arXiv:hep-th/0607085} \BibitemShut
  {NoStop}%
\bibitem [{\citenamefont {Henningson}\ and\ \citenamefont
  {Skenderis}(1998)}]{Henningson:1998gx}%
  \BibitemOpen
  \bibfield  {author} {\bibinfo {author} {\bibfnamefont {M.}~\bibnamefont
  {Henningson}}\ and\ \bibinfo {author} {\bibfnamefont {K.}~\bibnamefont
  {Skenderis}},\ }\href {\doibase 10.1088/1126-6708/1998/07/023} {\bibfield
  {journal} {\bibinfo  {journal} {JHEP}\ }\textbf {\bibinfo {volume} {07}},\
  \bibinfo {pages} {023} (\bibinfo {year} {1998})},\ \Eprint
  {http://arxiv.org/abs/hep-th/9806087} {arXiv:hep-th/9806087} \BibitemShut
  {NoStop}%
\bibitem [{\citenamefont {Hosseini}\ \emph {et~al.}(2017)\citenamefont
  {Hosseini}, \citenamefont {Hristov},\ and\ \citenamefont
  {Zaffaroni}}]{Hosseini:2017mds}%
  \BibitemOpen
  \bibfield  {author} {\bibinfo {author} {\bibfnamefont {S.~M.}\ \bibnamefont
  {Hosseini}}, \bibinfo {author} {\bibfnamefont {K.}~\bibnamefont {Hristov}}, \
  and\ \bibinfo {author} {\bibfnamefont {A.}~\bibnamefont {Zaffaroni}},\ }\href
  {\doibase 10.1007/JHEP07(2017)106} {\bibfield  {journal} {\bibinfo  {journal}
  {JHEP}\ }\textbf {\bibinfo {volume} {07}},\ \bibinfo {pages} {106} (\bibinfo
  {year} {2017})},\ \Eprint {http://arxiv.org/abs/1705.05383} {arXiv:1705.05383
  [hep-th]} \BibitemShut {NoStop}%
\bibitem [{\citenamefont {Cabo-Bizet}\ \emph
  {et~al.}(2019{\natexlab{b}})\citenamefont {Cabo-Bizet}, \citenamefont
  {Cassani}, \citenamefont {Martelli},\ and\ \citenamefont
  {Murthy}}]{Cabo-Bizet:2018ehj}%
  \BibitemOpen
  \bibfield  {author} {\bibinfo {author} {\bibfnamefont {A.}~\bibnamefont
  {Cabo-Bizet}}, \bibinfo {author} {\bibfnamefont {D.}~\bibnamefont {Cassani}},
  \bibinfo {author} {\bibfnamefont {D.}~\bibnamefont {Martelli}}, \ and\
  \bibinfo {author} {\bibfnamefont {S.}~\bibnamefont {Murthy}},\ }\href
  {\doibase 10.1007/JHEP10(2019)062} {\bibfield  {journal} {\bibinfo  {journal}
  {JHEP}\ }\textbf {\bibinfo {volume} {10}},\ \bibinfo {pages} {062} (\bibinfo
  {year} {2019}{\natexlab{b}})},\ \Eprint {http://arxiv.org/abs/1810.11442}
  {arXiv:1810.11442 [hep-th]} \BibitemShut {NoStop}%
\bibitem [{\citenamefont {Gaiotto}\ and\ \citenamefont
  {Maldacena}(2012)}]{Gaiotto:2009gz}%
  \BibitemOpen
  \bibfield  {author} {\bibinfo {author} {\bibfnamefont {D.}~\bibnamefont
  {Gaiotto}}\ and\ \bibinfo {author} {\bibfnamefont {J.}~\bibnamefont
  {Maldacena}},\ }\href {\doibase 10.1007/JHEP10(2012)189} {\bibfield
  {journal} {\bibinfo  {journal} {JHEP}\ }\textbf {\bibinfo {volume} {10}},\
  \bibinfo {pages} {189} (\bibinfo {year} {2012})},\ \Eprint
  {http://arxiv.org/abs/0904.4466} {arXiv:0904.4466 [hep-th]} \BibitemShut
  {NoStop}%
\bibitem [{\citenamefont {Maldacena}\ and\ \citenamefont
  {Nunez}(2001)}]{Maldacena:2000mw}%
  \BibitemOpen
  \bibfield  {author} {\bibinfo {author} {\bibfnamefont {J.~M.}\ \bibnamefont
  {Maldacena}}\ and\ \bibinfo {author} {\bibfnamefont {C.}~\bibnamefont
  {Nunez}},\ }\href {\doibase 10.1142/S0217751X01003937} {\bibfield  {journal}
  {\bibinfo  {journal} {Int. J. Mod. Phys. A}\ }\textbf {\bibinfo {volume}
  {16}},\ \bibinfo {pages} {822} (\bibinfo {year} {2001})},\ \Eprint
  {http://arxiv.org/abs/hep-th/0007018} {arXiv:hep-th/0007018} \BibitemShut
  {NoStop}%
\bibitem [{\citenamefont {Szepietowski}(2012)}]{Szepietowski:2012tb}%
  \BibitemOpen
  \bibfield  {author} {\bibinfo {author} {\bibfnamefont {P.}~\bibnamefont
  {Szepietowski}},\ }\href {\doibase 10.1007/JHEP12(2012)018} {\bibfield
  {journal} {\bibinfo  {journal} {JHEP}\ }\textbf {\bibinfo {volume} {12}},\
  \bibinfo {pages} {018} (\bibinfo {year} {2012})},\ \Eprint
  {http://arxiv.org/abs/1209.3025} {arXiv:1209.3025 [hep-th]} \BibitemShut
  {NoStop}%
\bibitem [{\citenamefont {Cassani}\ \emph {et~al.}(2020)\citenamefont
  {Cassani}, \citenamefont {Josse}, \citenamefont {Petrini},\ and\
  \citenamefont {Waldram}}]{Cassani:2020cod}%
  \BibitemOpen
  \bibfield  {author} {\bibinfo {author} {\bibfnamefont {D.}~\bibnamefont
  {Cassani}}, \bibinfo {author} {\bibfnamefont {G.}~\bibnamefont {Josse}},
  \bibinfo {author} {\bibfnamefont {M.}~\bibnamefont {Petrini}}, \ and\
  \bibinfo {author} {\bibfnamefont {D.}~\bibnamefont {Waldram}},\ }\href@noop
  {} {\  (\bibinfo {year} {2020})},\ \Eprint {http://arxiv.org/abs/2011.04775}
  {arXiv:2011.04775 [hep-th]} \BibitemShut {NoStop}%
\bibitem [{\citenamefont {Malek}\ and\ \citenamefont
  {Vall~Camell}(2020)}]{Malek:2020jsa}%
  \BibitemOpen
  \bibfield  {author} {\bibinfo {author} {\bibfnamefont {E.}~\bibnamefont
  {Malek}}\ and\ \bibinfo {author} {\bibfnamefont {V.}~\bibnamefont
  {Vall~Camell}},\ }\href@noop {} {\  (\bibinfo {year} {2020})},\ \Eprint
  {http://arxiv.org/abs/2012.15601} {arXiv:2012.15601 [hep-th]} \BibitemShut
  {NoStop}%
\bibitem [{\citenamefont {Hosseini}\ \emph {et~al.}(2018)\citenamefont
  {Hosseini}, \citenamefont {Hristov},\ and\ \citenamefont
  {Zaffaroni}}]{Hosseini:2018dob}%
  \BibitemOpen
  \bibfield  {author} {\bibinfo {author} {\bibfnamefont {S.~M.}\ \bibnamefont
  {Hosseini}}, \bibinfo {author} {\bibfnamefont {K.}~\bibnamefont {Hristov}}, \
  and\ \bibinfo {author} {\bibfnamefont {A.}~\bibnamefont {Zaffaroni}},\ }\href
  {\doibase 10.1007/JHEP05(2018)121} {\bibfield  {journal} {\bibinfo  {journal}
  {JHEP}\ }\textbf {\bibinfo {volume} {05}},\ \bibinfo {pages} {121} (\bibinfo
  {year} {2018})},\ \Eprint {http://arxiv.org/abs/1803.07568} {arXiv:1803.07568
  [hep-th]} \BibitemShut {NoStop}%
\bibitem [{\citenamefont {Intriligator}\ and\ \citenamefont
  {Wecht}(2003)}]{Intriligator:2003jj}%
  \BibitemOpen
  \bibfield  {author} {\bibinfo {author} {\bibfnamefont {K.~A.}\ \bibnamefont
  {Intriligator}}\ and\ \bibinfo {author} {\bibfnamefont {B.}~\bibnamefont
  {Wecht}},\ }\href {\doibase 10.1016/S0550-3213(03)00459-0} {\bibfield
  {journal} {\bibinfo  {journal} {Nucl. Phys. B}\ }\textbf {\bibinfo {volume}
  {667}},\ \bibinfo {pages} {183} (\bibinfo {year} {2003})},\ \Eprint
  {http://arxiv.org/abs/hep-th/0304128} {arXiv:hep-th/0304128} \BibitemShut
  {NoStop}%
\bibitem [{\citenamefont {Lanir}\ \emph {et~al.}(2020)\citenamefont {Lanir},
  \citenamefont {Nedelin},\ and\ \citenamefont {Sela}}]{Lanir:2019abx}%
  \BibitemOpen
  \bibfield  {author} {\bibinfo {author} {\bibfnamefont {A.}~\bibnamefont
  {Lanir}}, \bibinfo {author} {\bibfnamefont {A.}~\bibnamefont {Nedelin}}, \
  and\ \bibinfo {author} {\bibfnamefont {O.}~\bibnamefont {Sela}},\ }\href
  {\doibase 10.1007/JHEP04(2020)091} {\bibfield  {journal} {\bibinfo  {journal}
  {JHEP}\ }\textbf {\bibinfo {volume} {04}},\ \bibinfo {pages} {091} (\bibinfo
  {year} {2020})},\ \Eprint {http://arxiv.org/abs/1908.01737} {arXiv:1908.01737
  [hep-th]} \BibitemShut {NoStop}%
\bibitem [{\citenamefont {Cabo-Bizet}\ \emph {et~al.}(2020)\citenamefont
  {Cabo-Bizet}, \citenamefont {Cassani}, \citenamefont {Martelli},\ and\
  \citenamefont {Murthy}}]{Cabo-Bizet:2020nkr}%
  \BibitemOpen
  \bibfield  {author} {\bibinfo {author} {\bibfnamefont {A.}~\bibnamefont
  {Cabo-Bizet}}, \bibinfo {author} {\bibfnamefont {D.}~\bibnamefont {Cassani}},
  \bibinfo {author} {\bibfnamefont {D.}~\bibnamefont {Martelli}}, \ and\
  \bibinfo {author} {\bibfnamefont {S.}~\bibnamefont {Murthy}},\ }\href@noop {}
  {\  (\bibinfo {year} {2020})},\ \Eprint {http://arxiv.org/abs/2005.10654}
  {arXiv:2005.10654 [hep-th]} \BibitemShut {NoStop}%
\bibitem [{\citenamefont {Amariti}\ \emph {et~al.}(2019)\citenamefont
  {Amariti}, \citenamefont {Garozzo},\ and\ \citenamefont
  {Lo~Monaco}}]{Amariti:2019mgp}%
  \BibitemOpen
  \bibfield  {author} {\bibinfo {author} {\bibfnamefont {A.}~\bibnamefont
  {Amariti}}, \bibinfo {author} {\bibfnamefont {I.}~\bibnamefont {Garozzo}}, \
  and\ \bibinfo {author} {\bibfnamefont {G.}~\bibnamefont {Lo~Monaco}},\
  }\href@noop {} {\  (\bibinfo {year} {2019})},\ \Eprint
  {http://arxiv.org/abs/1904.10009} {arXiv:1904.10009 [hep-th]} \BibitemShut
  {NoStop}%
\bibitem [{\citenamefont {Anselmi}\ \emph {et~al.}(1998)\citenamefont
  {Anselmi}, \citenamefont {Freedman}, \citenamefont {Grisaru},\ and\
  \citenamefont {Johansen}}]{Anselmi:1997am}%
  \BibitemOpen
  \bibfield  {author} {\bibinfo {author} {\bibfnamefont {D.}~\bibnamefont
  {Anselmi}}, \bibinfo {author} {\bibfnamefont {D.}~\bibnamefont {Freedman}},
  \bibinfo {author} {\bibfnamefont {M.~T.}\ \bibnamefont {Grisaru}}, \ and\
  \bibinfo {author} {\bibfnamefont {A.}~\bibnamefont {Johansen}},\ }\href
  {\doibase 10.1016/S0550-3213(98)00278-8} {\bibfield  {journal} {\bibinfo
  {journal} {Nucl. Phys. B}\ }\textbf {\bibinfo {volume} {526}},\ \bibinfo
  {pages} {543} (\bibinfo {year} {1998})},\ \Eprint
  {http://arxiv.org/abs/hep-th/9708042} {arXiv:hep-th/9708042} \BibitemShut
  {NoStop}%
\bibitem [{\citenamefont {Shapere}\ and\ \citenamefont
  {Tachikawa}(2008)}]{Shapere:2008zf}%
  \BibitemOpen
  \bibfield  {author} {\bibinfo {author} {\bibfnamefont {A.~D.}\ \bibnamefont
  {Shapere}}\ and\ \bibinfo {author} {\bibfnamefont {Y.}~\bibnamefont
  {Tachikawa}},\ }\href {\doibase 10.1088/1126-6708/2008/09/109} {\bibfield
  {journal} {\bibinfo  {journal} {JHEP}\ }\textbf {\bibinfo {volume} {09}},\
  \bibinfo {pages} {109} (\bibinfo {year} {2008})},\ \Eprint
  {http://arxiv.org/abs/0804.1957} {arXiv:0804.1957 [hep-th]} \BibitemShut
  {NoStop}%
\bibitem [{\citenamefont {Hosseini}\ \emph
  {et~al.}(2019{\natexlab{a}})\citenamefont {Hosseini}, \citenamefont
  {Hristov},\ and\ \citenamefont {Zaffaroni}}]{Hosseini:2019lkt}%
  \BibitemOpen
  \bibfield  {author} {\bibinfo {author} {\bibfnamefont {S.~M.}\ \bibnamefont
  {Hosseini}}, \bibinfo {author} {\bibfnamefont {K.}~\bibnamefont {Hristov}}, \
  and\ \bibinfo {author} {\bibfnamefont {A.}~\bibnamefont {Zaffaroni}},\ }\href
  {\doibase 10.1007/JHEP11(2019)090} {\bibfield  {journal} {\bibinfo  {journal}
  {JHEP}\ }\textbf {\bibinfo {volume} {11}},\ \bibinfo {pages} {090} (\bibinfo
  {year} {2019}{\natexlab{a}})},\ \Eprint {http://arxiv.org/abs/1909.08000}
  {arXiv:1909.08000 [hep-th]} \BibitemShut {NoStop}%
\bibitem [{\citenamefont {Hosseini}\ \emph {et~al.}(2020)\citenamefont
  {Hosseini}, \citenamefont {Hristov}, \citenamefont {Tachikawa},\ and\
  \citenamefont {Zaffaroni}}]{Hosseini:2020vgl}%
  \BibitemOpen
  \bibfield  {author} {\bibinfo {author} {\bibfnamefont {S.~M.}\ \bibnamefont
  {Hosseini}}, \bibinfo {author} {\bibfnamefont {K.}~\bibnamefont {Hristov}},
  \bibinfo {author} {\bibfnamefont {Y.}~\bibnamefont {Tachikawa}}, \ and\
  \bibinfo {author} {\bibfnamefont {A.}~\bibnamefont {Zaffaroni}},\ }\href
  {\doibase 10.1007/JHEP09(2020)167} {\bibfield  {journal} {\bibinfo  {journal}
  {JHEP}\ }\textbf {\bibinfo {volume} {09}},\ \bibinfo {pages} {167} (\bibinfo
  {year} {2020})},\ \Eprint {http://arxiv.org/abs/2006.08629} {arXiv:2006.08629
  [hep-th]} \BibitemShut {NoStop}%
\bibitem [{\citenamefont {Benini}\ and\ \citenamefont
  {Bobev}(2013)}]{Benini:2013cda}%
  \BibitemOpen
  \bibfield  {author} {\bibinfo {author} {\bibfnamefont {F.}~\bibnamefont
  {Benini}}\ and\ \bibinfo {author} {\bibfnamefont {N.}~\bibnamefont {Bobev}},\
  }\href {\doibase 10.1007/JHEP06(2013)005} {\bibfield  {journal} {\bibinfo
  {journal} {JHEP}\ }\textbf {\bibinfo {volume} {06}},\ \bibinfo {pages} {005}
  (\bibinfo {year} {2013})},\ \Eprint {http://arxiv.org/abs/1302.4451}
  {arXiv:1302.4451 [hep-th]} \BibitemShut {NoStop}%
\bibitem [{\citenamefont {Bah}\ \emph {et~al.}(2020)\citenamefont {Bah},
  \citenamefont {Bonetti}, \citenamefont {Minasian},\ and\ \citenamefont
  {Nardoni}}]{Bah:2019rgq}%
  \BibitemOpen
  \bibfield  {author} {\bibinfo {author} {\bibfnamefont {I.}~\bibnamefont
  {Bah}}, \bibinfo {author} {\bibfnamefont {F.}~\bibnamefont {Bonetti}},
  \bibinfo {author} {\bibfnamefont {R.}~\bibnamefont {Minasian}}, \ and\
  \bibinfo {author} {\bibfnamefont {E.}~\bibnamefont {Nardoni}},\ }\href
  {\doibase 10.1007/JHEP01(2020)125} {\bibfield  {journal} {\bibinfo  {journal}
  {JHEP}\ }\textbf {\bibinfo {volume} {01}},\ \bibinfo {pages} {125} (\bibinfo
  {year} {2020})},\ \Eprint {http://arxiv.org/abs/1910.04166} {arXiv:1910.04166
  [hep-th]} \BibitemShut {NoStop}%
\bibitem [{\citenamefont {Cvetic}\ \emph {et~al.}(2000)\citenamefont {Cvetic},
  \citenamefont {Lu},\ and\ \citenamefont {Pope}}]{Cvetic:1999au}%
  \BibitemOpen
  \bibfield  {author} {\bibinfo {author} {\bibfnamefont {M.}~\bibnamefont
  {Cvetic}}, \bibinfo {author} {\bibfnamefont {H.}~\bibnamefont {Lu}}, \ and\
  \bibinfo {author} {\bibfnamefont {C.}~\bibnamefont {Pope}},\ }\href {\doibase
  10.1016/S0550-3213(99)00828-7} {\bibfield  {journal} {\bibinfo  {journal}
  {Nucl. Phys. B}\ }\textbf {\bibinfo {volume} {574}},\ \bibinfo {pages} {761}
  (\bibinfo {year} {2000})},\ \Eprint {http://arxiv.org/abs/hep-th/9910252}
  {arXiv:hep-th/9910252} \BibitemShut {NoStop}%
\bibitem [{\citenamefont {Kostelecky}\ and\ \citenamefont
  {Perry}(1996)}]{Kostelecky:1995ei}%
  \BibitemOpen
  \bibfield  {author} {\bibinfo {author} {\bibfnamefont {V.}~\bibnamefont
  {Kostelecky}}\ and\ \bibinfo {author} {\bibfnamefont {M.~J.}\ \bibnamefont
  {Perry}},\ }\href {\doibase 10.1016/0370-2693(95)01607-4} {\bibfield
  {journal} {\bibinfo  {journal} {Phys. Lett. B}\ }\textbf {\bibinfo {volume}
  {371}},\ \bibinfo {pages} {191} (\bibinfo {year} {1996})},\ \Eprint
  {http://arxiv.org/abs/hep-th/9512222} {arXiv:hep-th/9512222} \BibitemShut
  {NoStop}%
\bibitem [{\citenamefont {Cvetic}\ \emph {et~al.}(2005)\citenamefont {Cvetic},
  \citenamefont {Gibbons}, \citenamefont {Lu},\ and\ \citenamefont
  {Pope}}]{Cvetic:2005zi}%
  \BibitemOpen
  \bibfield  {author} {\bibinfo {author} {\bibfnamefont {M.}~\bibnamefont
  {Cvetic}}, \bibinfo {author} {\bibfnamefont {G.}~\bibnamefont {Gibbons}},
  \bibinfo {author} {\bibfnamefont {H.}~\bibnamefont {Lu}}, \ and\ \bibinfo
  {author} {\bibfnamefont {C.}~\bibnamefont {Pope}},\ }\href@noop {} {\
  (\bibinfo {year} {2005})},\ \Eprint {http://arxiv.org/abs/hep-th/0504080}
  {arXiv:hep-th/0504080} \BibitemShut {NoStop}%
\bibitem [{\citenamefont {Hristov}\ \emph
  {et~al.}(2019{\natexlab{a}})\citenamefont {Hristov}, \citenamefont
  {Katmadas},\ and\ \citenamefont {Toldo}}]{Hristov:2019mqp}%
  \BibitemOpen
  \bibfield  {author} {\bibinfo {author} {\bibfnamefont {K.}~\bibnamefont
  {Hristov}}, \bibinfo {author} {\bibfnamefont {S.}~\bibnamefont {Katmadas}}, \
  and\ \bibinfo {author} {\bibfnamefont {C.}~\bibnamefont {Toldo}},\ }\href
  {\doibase 10.1103/PhysRevD.100.066016} {\bibfield  {journal} {\bibinfo
  {journal} {Phys. Rev. D}\ }\textbf {\bibinfo {volume} {100}},\ \bibinfo
  {pages} {066016} (\bibinfo {year} {2019}{\natexlab{a}})},\ \Eprint
  {http://arxiv.org/abs/1907.05192} {arXiv:1907.05192 [hep-th]} \BibitemShut
  {NoStop}%
\bibitem [{\citenamefont {Choi}\ \emph {et~al.}(2020)\citenamefont {Choi},
  \citenamefont {Hwang}, \citenamefont {Kim},\ and\ \citenamefont
  {Nahmgoong}}]{Choi:2018fdc}%
  \BibitemOpen
  \bibfield  {author} {\bibinfo {author} {\bibfnamefont {S.}~\bibnamefont
  {Choi}}, \bibinfo {author} {\bibfnamefont {C.}~\bibnamefont {Hwang}},
  \bibinfo {author} {\bibfnamefont {S.}~\bibnamefont {Kim}}, \ and\ \bibinfo
  {author} {\bibfnamefont {J.}~\bibnamefont {Nahmgoong}},\ }\href {\doibase
  10.3938/jkps.76.101} {\bibfield  {journal} {\bibinfo  {journal} {J. Korean
  Phys. Soc.}\ }\textbf {\bibinfo {volume} {76}},\ \bibinfo {pages} {101}
  (\bibinfo {year} {2020})},\ \Eprint {http://arxiv.org/abs/1811.02158}
  {arXiv:1811.02158 [hep-th]} \BibitemShut {NoStop}%
\bibitem [{\citenamefont {Hosseini}\ \emph
  {et~al.}(2019{\natexlab{b}})\citenamefont {Hosseini}, \citenamefont
  {Hristov},\ and\ \citenamefont {Zaffaroni}}]{Hosseini:2019iad}%
  \BibitemOpen
  \bibfield  {author} {\bibinfo {author} {\bibfnamefont {S.~M.}\ \bibnamefont
  {Hosseini}}, \bibinfo {author} {\bibfnamefont {K.}~\bibnamefont {Hristov}}, \
  and\ \bibinfo {author} {\bibfnamefont {A.}~\bibnamefont {Zaffaroni}},\ }\href
  {\doibase 10.1007/JHEP12(2019)168} {\bibfield  {journal} {\bibinfo  {journal}
  {JHEP}\ }\textbf {\bibinfo {volume} {12}},\ \bibinfo {pages} {168} (\bibinfo
  {year} {2019}{\natexlab{b}})},\ \Eprint {http://arxiv.org/abs/1909.10550}
  {arXiv:1909.10550 [hep-th]} \BibitemShut {NoStop}%
\bibitem [{\citenamefont {Jafferis}(2012)}]{Jafferis:2010un}%
  \BibitemOpen
  \bibfield  {author} {\bibinfo {author} {\bibfnamefont {D.~L.}\ \bibnamefont
  {Jafferis}},\ }\href {\doibase 10.1007/JHEP05(2012)159} {\bibfield  {journal}
  {\bibinfo  {journal} {JHEP}\ }\textbf {\bibinfo {volume} {05}},\ \bibinfo
  {pages} {159} (\bibinfo {year} {2012})},\ \Eprint
  {http://arxiv.org/abs/1012.3210} {arXiv:1012.3210 [hep-th]} \BibitemShut
  {NoStop}%
\bibitem [{\citenamefont {Pasquetti}(2012)}]{Pasquetti:2011fj}%
  \BibitemOpen
  \bibfield  {author} {\bibinfo {author} {\bibfnamefont {S.}~\bibnamefont
  {Pasquetti}},\ }\href {\doibase 10.1007/JHEP04(2012)120} {\bibfield
  {journal} {\bibinfo  {journal} {JHEP}\ }\textbf {\bibinfo {volume} {04}},\
  \bibinfo {pages} {120} (\bibinfo {year} {2012})},\ \Eprint
  {http://arxiv.org/abs/1111.6905} {arXiv:1111.6905 [hep-th]} \BibitemShut
  {NoStop}%
\bibitem [{\citenamefont {Beem}\ \emph {et~al.}(2014)\citenamefont {Beem},
  \citenamefont {Dimofte},\ and\ \citenamefont {Pasquetti}}]{Beem:2012mb}%
  \BibitemOpen
  \bibfield  {author} {\bibinfo {author} {\bibfnamefont {C.}~\bibnamefont
  {Beem}}, \bibinfo {author} {\bibfnamefont {T.}~\bibnamefont {Dimofte}}, \
  and\ \bibinfo {author} {\bibfnamefont {S.}~\bibnamefont {Pasquetti}},\ }\href
  {\doibase 10.1007/JHEP12(2014)177} {\bibfield  {journal} {\bibinfo  {journal}
  {JHEP}\ }\textbf {\bibinfo {volume} {12}},\ \bibinfo {pages} {177} (\bibinfo
  {year} {2014})},\ \Eprint {http://arxiv.org/abs/1211.1986} {arXiv:1211.1986
  [hep-th]} \BibitemShut {NoStop}%
\bibitem [{\citenamefont {Choi}\ \emph {et~al.}(2019)\citenamefont {Choi},
  \citenamefont {Hwang},\ and\ \citenamefont {Kim}}]{Choi:2019zpz}%
  \BibitemOpen
  \bibfield  {author} {\bibinfo {author} {\bibfnamefont {S.}~\bibnamefont
  {Choi}}, \bibinfo {author} {\bibfnamefont {C.}~\bibnamefont {Hwang}}, \ and\
  \bibinfo {author} {\bibfnamefont {S.}~\bibnamefont {Kim}},\ }\href@noop {} {\
   (\bibinfo {year} {2019})},\ \Eprint {http://arxiv.org/abs/1908.02470}
  {arXiv:1908.02470 [hep-th]} \BibitemShut {NoStop}%
\bibitem [{\citenamefont {Choi}\ and\ \citenamefont
  {Hwang}(2020)}]{Choi:2019dfu}%
  \BibitemOpen
  \bibfield  {author} {\bibinfo {author} {\bibfnamefont {S.}~\bibnamefont
  {Choi}}\ and\ \bibinfo {author} {\bibfnamefont {C.}~\bibnamefont {Hwang}},\
  }\href {\doibase 10.1007/JHEP03(2020)068} {\bibfield  {journal} {\bibinfo
  {journal} {JHEP}\ }\textbf {\bibinfo {volume} {03}},\ \bibinfo {pages} {068}
  (\bibinfo {year} {2020})},\ \Eprint {http://arxiv.org/abs/1911.01448}
  {arXiv:1911.01448 [hep-th]} \BibitemShut {NoStop}%
\bibitem [{\citenamefont {Porrati}\ and\ \citenamefont
  {Zaffaroni}(1997)}]{Porrati:1996xi}%
  \BibitemOpen
  \bibfield  {author} {\bibinfo {author} {\bibfnamefont {M.}~\bibnamefont
  {Porrati}}\ and\ \bibinfo {author} {\bibfnamefont {A.}~\bibnamefont
  {Zaffaroni}},\ }\href {\doibase 10.1016/S0550-3213(97)00061-8} {\bibfield
  {journal} {\bibinfo  {journal} {Nucl. Phys. B}\ }\textbf {\bibinfo {volume}
  {490}},\ \bibinfo {pages} {107} (\bibinfo {year} {1997})},\ \Eprint
  {http://arxiv.org/abs/hep-th/9611201} {arXiv:hep-th/9611201} \BibitemShut
  {NoStop}%
\bibitem [{\citenamefont {Mekareeya}(2015)}]{Mekareeya:2015bla}%
  \BibitemOpen
  \bibfield  {author} {\bibinfo {author} {\bibfnamefont {N.}~\bibnamefont
  {Mekareeya}},\ }\href {\doibase 10.1007/JHEP12(2015)174} {\bibfield
  {journal} {\bibinfo  {journal} {JHEP}\ }\textbf {\bibinfo {volume} {12}},\
  \bibinfo {pages} {174} (\bibinfo {year} {2015})},\ \Eprint
  {http://arxiv.org/abs/1508.06813} {arXiv:1508.06813 [hep-th]} \BibitemShut
  {NoStop}%
\bibitem [{\citenamefont {Herzog}\ \emph {et~al.}(2011)\citenamefont {Herzog},
  \citenamefont {Klebanov}, \citenamefont {Pufu},\ and\ \citenamefont
  {Tesileanu}}]{Herzog:2010hf}%
  \BibitemOpen
  \bibfield  {author} {\bibinfo {author} {\bibfnamefont {C.~P.}\ \bibnamefont
  {Herzog}}, \bibinfo {author} {\bibfnamefont {I.~R.}\ \bibnamefont
  {Klebanov}}, \bibinfo {author} {\bibfnamefont {S.~S.}\ \bibnamefont {Pufu}},
  \ and\ \bibinfo {author} {\bibfnamefont {T.}~\bibnamefont {Tesileanu}},\
  }\href {\doibase 10.1103/PhysRevD.83.046001} {\bibfield  {journal} {\bibinfo
  {journal} {Phys. Rev. D}\ }\textbf {\bibinfo {volume} {83}},\ \bibinfo
  {pages} {046001} (\bibinfo {year} {2011})},\ \Eprint
  {http://arxiv.org/abs/1011.5487} {arXiv:1011.5487 [hep-th]} \BibitemShut
  {NoStop}%
\bibitem [{\citenamefont {Jafferis}\ \emph {et~al.}(2011)\citenamefont
  {Jafferis}, \citenamefont {Klebanov}, \citenamefont {Pufu},\ and\
  \citenamefont {Safdi}}]{Jafferis:2011zi}%
  \BibitemOpen
  \bibfield  {author} {\bibinfo {author} {\bibfnamefont {D.~L.}\ \bibnamefont
  {Jafferis}}, \bibinfo {author} {\bibfnamefont {I.~R.}\ \bibnamefont
  {Klebanov}}, \bibinfo {author} {\bibfnamefont {S.~S.}\ \bibnamefont {Pufu}},
  \ and\ \bibinfo {author} {\bibfnamefont {B.~R.}\ \bibnamefont {Safdi}},\
  }\href {\doibase 10.1007/JHEP06(2011)102} {\bibfield  {journal} {\bibinfo
  {journal} {JHEP}\ }\textbf {\bibinfo {volume} {06}},\ \bibinfo {pages} {102}
  (\bibinfo {year} {2011})},\ \Eprint {http://arxiv.org/abs/1103.1181}
  {arXiv:1103.1181 [hep-th]} \BibitemShut {NoStop}%
\bibitem [{\citenamefont {Crichigno}\ \emph {et~al.}(2013)\citenamefont
  {Crichigno}, \citenamefont {Herzog},\ and\ \citenamefont
  {Jain}}]{Crichigno:2012sk}%
  \BibitemOpen
  \bibfield  {author} {\bibinfo {author} {\bibfnamefont {P.~M.}\ \bibnamefont
  {Crichigno}}, \bibinfo {author} {\bibfnamefont {C.~P.}\ \bibnamefont
  {Herzog}}, \ and\ \bibinfo {author} {\bibfnamefont {D.}~\bibnamefont
  {Jain}},\ }\href {\doibase 10.1007/JHEP03(2013)039} {\bibfield  {journal}
  {\bibinfo  {journal} {JHEP}\ }\textbf {\bibinfo {volume} {03}},\ \bibinfo
  {pages} {039} (\bibinfo {year} {2013})},\ \Eprint
  {http://arxiv.org/abs/1211.1388} {arXiv:1211.1388 [hep-th]} \BibitemShut
  {NoStop}%
\bibitem [{\citenamefont {Hosseini}\ and\ \citenamefont
  {Mekareeya}(2016)}]{Hosseini:2016ume}%
  \BibitemOpen
  \bibfield  {author} {\bibinfo {author} {\bibfnamefont {S.~M.}\ \bibnamefont
  {Hosseini}}\ and\ \bibinfo {author} {\bibfnamefont {N.}~\bibnamefont
  {Mekareeya}},\ }\href {\doibase 10.1007/JHEP08(2016)089} {\bibfield
  {journal} {\bibinfo  {journal} {JHEP}\ }\textbf {\bibinfo {volume} {08}},\
  \bibinfo {pages} {089} (\bibinfo {year} {2016})},\ \Eprint
  {http://arxiv.org/abs/1604.03397} {arXiv:1604.03397 [hep-th]} \BibitemShut
  {NoStop}%
\bibitem [{\citenamefont {Martelli}\ \emph {et~al.}(2008)\citenamefont
  {Martelli}, \citenamefont {Sparks},\ and\ \citenamefont
  {Yau}}]{Martelli:2006yb}%
  \BibitemOpen
  \bibfield  {author} {\bibinfo {author} {\bibfnamefont {D.}~\bibnamefont
  {Martelli}}, \bibinfo {author} {\bibfnamefont {J.}~\bibnamefont {Sparks}}, \
  and\ \bibinfo {author} {\bibfnamefont {S.-T.}\ \bibnamefont {Yau}},\ }\href
  {\doibase 10.1007/s00220-008-0479-4} {\bibfield  {journal} {\bibinfo
  {journal} {Commun. Math. Phys.}\ }\textbf {\bibinfo {volume} {280}},\
  \bibinfo {pages} {611} (\bibinfo {year} {2008})},\ \Eprint
  {http://arxiv.org/abs/hep-th/0603021} {arXiv:hep-th/0603021} \BibitemShut
  {NoStop}%
\bibitem [{\citenamefont {Imamura}\ and\ \citenamefont
  {Kimura}(2008{\natexlab{a}})}]{Imamura:2008nn}%
  \BibitemOpen
  \bibfield  {author} {\bibinfo {author} {\bibfnamefont {Y.}~\bibnamefont
  {Imamura}}\ and\ \bibinfo {author} {\bibfnamefont {K.}~\bibnamefont
  {Kimura}},\ }\href {\doibase 10.1143/PTP.120.509} {\bibfield  {journal}
  {\bibinfo  {journal} {Prog. Theor. Phys.}\ }\textbf {\bibinfo {volume}
  {120}},\ \bibinfo {pages} {509} (\bibinfo {year} {2008}{\natexlab{a}})},\
  \Eprint {http://arxiv.org/abs/0806.3727} {arXiv:0806.3727 [hep-th]}
  \BibitemShut {NoStop}%
\bibitem [{\citenamefont {Amariti}\ \emph {et~al.}(2012)\citenamefont
  {Amariti}, \citenamefont {Klare},\ and\ \citenamefont
  {Siani}}]{Amariti:2011uw}%
  \BibitemOpen
  \bibfield  {author} {\bibinfo {author} {\bibfnamefont {A.}~\bibnamefont
  {Amariti}}, \bibinfo {author} {\bibfnamefont {C.}~\bibnamefont {Klare}}, \
  and\ \bibinfo {author} {\bibfnamefont {M.}~\bibnamefont {Siani}},\ }\href
  {\doibase 10.1007/JHEP10(2012)019} {\bibfield  {journal} {\bibinfo  {journal}
  {JHEP}\ }\textbf {\bibinfo {volume} {10}},\ \bibinfo {pages} {019} (\bibinfo
  {year} {2012})},\ \Eprint {http://arxiv.org/abs/1111.1723} {arXiv:1111.1723
  [hep-th]} \BibitemShut {NoStop}%
\bibitem [{\citenamefont {Gaiotto}\ and\ \citenamefont
  {Witten}(2010)}]{Gaiotto:2008sd}%
  \BibitemOpen
  \bibfield  {author} {\bibinfo {author} {\bibfnamefont {D.}~\bibnamefont
  {Gaiotto}}\ and\ \bibinfo {author} {\bibfnamefont {E.}~\bibnamefont
  {Witten}},\ }\href {\doibase 10.1007/JHEP06(2010)097} {\bibfield  {journal}
  {\bibinfo  {journal} {JHEP}\ }\textbf {\bibinfo {volume} {06}},\ \bibinfo
  {pages} {097} (\bibinfo {year} {2010})},\ \Eprint
  {http://arxiv.org/abs/0804.2907} {arXiv:0804.2907 [hep-th]} \BibitemShut
  {NoStop}%
\bibitem [{\citenamefont {Imamura}\ and\ \citenamefont
  {Kimura}(2008{\natexlab{b}})}]{Imamura:2008dt}%
  \BibitemOpen
  \bibfield  {author} {\bibinfo {author} {\bibfnamefont {Y.}~\bibnamefont
  {Imamura}}\ and\ \bibinfo {author} {\bibfnamefont {K.}~\bibnamefont
  {Kimura}},\ }\href {\doibase 10.1088/1126-6708/2008/10/040} {\bibfield
  {journal} {\bibinfo  {journal} {JHEP}\ }\textbf {\bibinfo {volume} {10}},\
  \bibinfo {pages} {040} (\bibinfo {year} {2008}{\natexlab{b}})},\ \Eprint
  {http://arxiv.org/abs/0807.2144} {arXiv:0807.2144 [hep-th]} \BibitemShut
  {NoStop}%
\bibitem [{\citenamefont {Hosomichi}\ \emph {et~al.}(2008)\citenamefont
  {Hosomichi}, \citenamefont {Lee}, \citenamefont {Lee}, \citenamefont {Lee},\
  and\ \citenamefont {Park}}]{Hosomichi:2008jd}%
  \BibitemOpen
  \bibfield  {author} {\bibinfo {author} {\bibfnamefont {K.}~\bibnamefont
  {Hosomichi}}, \bibinfo {author} {\bibfnamefont {K.-M.}\ \bibnamefont {Lee}},
  \bibinfo {author} {\bibfnamefont {S.}~\bibnamefont {Lee}}, \bibinfo {author}
  {\bibfnamefont {S.}~\bibnamefont {Lee}}, \ and\ \bibinfo {author}
  {\bibfnamefont {J.}~\bibnamefont {Park}},\ }\href {\doibase
  10.1088/1126-6708/2008/07/091} {\bibfield  {journal} {\bibinfo  {journal}
  {JHEP}\ }\textbf {\bibinfo {volume} {07}},\ \bibinfo {pages} {091} (\bibinfo
  {year} {2008})},\ \Eprint {http://arxiv.org/abs/0805.3662} {arXiv:0805.3662
  [hep-th]} \BibitemShut {NoStop}%
\bibitem [{\citenamefont {Gaiotto}\ and\ \citenamefont
  {Witten}(2009)}]{Gaiotto:2008sa}%
  \BibitemOpen
  \bibfield  {author} {\bibinfo {author} {\bibfnamefont {D.}~\bibnamefont
  {Gaiotto}}\ and\ \bibinfo {author} {\bibfnamefont {E.}~\bibnamefont
  {Witten}},\ }\href {\doibase 10.1007/s10955-009-9687-3} {\bibfield  {journal}
  {\bibinfo  {journal} {J. Statist. Phys.}\ }\textbf {\bibinfo {volume}
  {135}},\ \bibinfo {pages} {789} (\bibinfo {year} {2009})},\ \Eprint
  {http://arxiv.org/abs/0804.2902} {arXiv:0804.2902 [hep-th]} \BibitemShut
  {NoStop}%
\bibitem [{\citenamefont {Assel}\ \emph {et~al.}(2011)\citenamefont {Assel},
  \citenamefont {Bachas}, \citenamefont {Estes},\ and\ \citenamefont
  {Gomis}}]{Assel:2011xz}%
  \BibitemOpen
  \bibfield  {author} {\bibinfo {author} {\bibfnamefont {B.}~\bibnamefont
  {Assel}}, \bibinfo {author} {\bibfnamefont {C.}~\bibnamefont {Bachas}},
  \bibinfo {author} {\bibfnamefont {J.}~\bibnamefont {Estes}}, \ and\ \bibinfo
  {author} {\bibfnamefont {J.}~\bibnamefont {Gomis}},\ }\href {\doibase
  10.1007/JHEP08(2011)087} {\bibfield  {journal} {\bibinfo  {journal} {JHEP}\
  }\textbf {\bibinfo {volume} {08}},\ \bibinfo {pages} {087} (\bibinfo {year}
  {2011})},\ \Eprint {http://arxiv.org/abs/1106.4253} {arXiv:1106.4253
  [hep-th]} \BibitemShut {NoStop}%
\bibitem [{\citenamefont {Assel}\ \emph {et~al.}(2012)\citenamefont {Assel},
  \citenamefont {Bachas}, \citenamefont {Estes},\ and\ \citenamefont
  {Gomis}}]{Assel:2012cj}%
  \BibitemOpen
  \bibfield  {author} {\bibinfo {author} {\bibfnamefont {B.}~\bibnamefont
  {Assel}}, \bibinfo {author} {\bibfnamefont {C.}~\bibnamefont {Bachas}},
  \bibinfo {author} {\bibfnamefont {J.}~\bibnamefont {Estes}}, \ and\ \bibinfo
  {author} {\bibfnamefont {J.}~\bibnamefont {Gomis}},\ }\href {\doibase
  10.1007/JHEP12(2012)044} {\bibfield  {journal} {\bibinfo  {journal} {JHEP}\
  }\textbf {\bibinfo {volume} {12}},\ \bibinfo {pages} {044} (\bibinfo {year}
  {2012})},\ \Eprint {http://arxiv.org/abs/1210.2590} {arXiv:1210.2590
  [hep-th]} \BibitemShut {NoStop}%
\bibitem [{\citenamefont {Coccia}(2020)}]{Coccia:2020cku}%
  \BibitemOpen
  \bibfield  {author} {\bibinfo {author} {\bibfnamefont {L.}~\bibnamefont
  {Coccia}},\ }\href@noop {} {\  (\bibinfo {year} {2020})},\ \Eprint
  {http://arxiv.org/abs/2006.06578} {arXiv:2006.06578 [hep-th]} \BibitemShut
  {NoStop}%
\bibitem [{\citenamefont {Coccia}\ and\ \citenamefont
  {Uhlemann}(2020)}]{Coccia:2020wtk}%
  \BibitemOpen
  \bibfield  {author} {\bibinfo {author} {\bibfnamefont {L.}~\bibnamefont
  {Coccia}}\ and\ \bibinfo {author} {\bibfnamefont {C.~F.}\ \bibnamefont
  {Uhlemann}},\ }\href@noop {} {\  (\bibinfo {year} {2020})},\ \Eprint
  {http://arxiv.org/abs/2011.10050} {arXiv:2011.10050 [hep-th]} \BibitemShut
  {NoStop}%
\bibitem [{\citenamefont {Cacciatori}\ and\ \citenamefont
  {Klemm}(2010)}]{Cacciatori:2009iz}%
  \BibitemOpen
  \bibfield  {author} {\bibinfo {author} {\bibfnamefont {S.~L.}\ \bibnamefont
  {Cacciatori}}\ and\ \bibinfo {author} {\bibfnamefont {D.}~\bibnamefont
  {Klemm}},\ }\href {\doibase 10.1007/JHEP01(2010)085} {\bibfield  {journal}
  {\bibinfo  {journal} {JHEP}\ }\textbf {\bibinfo {volume} {01}},\ \bibinfo
  {pages} {085} (\bibinfo {year} {2010})},\ \Eprint
  {http://arxiv.org/abs/0911.4926} {arXiv:0911.4926 [hep-th]} \BibitemShut
  {NoStop}%
\bibitem [{\citenamefont {Katmadas}(2014)}]{Katmadas:2014faa}%
  \BibitemOpen
  \bibfield  {author} {\bibinfo {author} {\bibfnamefont {S.}~\bibnamefont
  {Katmadas}},\ }\href {\doibase 10.1007/JHEP09(2014)027} {\bibfield  {journal}
  {\bibinfo  {journal} {JHEP}\ }\textbf {\bibinfo {volume} {09}},\ \bibinfo
  {pages} {027} (\bibinfo {year} {2014})},\ \Eprint
  {http://arxiv.org/abs/1405.4901} {arXiv:1405.4901 [hep-th]} \BibitemShut
  {NoStop}%
\bibitem [{\citenamefont {Halmagyi}(2015)}]{Halmagyi:2014qza}%
  \BibitemOpen
  \bibfield  {author} {\bibinfo {author} {\bibfnamefont {N.}~\bibnamefont
  {Halmagyi}},\ }\href {\doibase 10.1007/JHEP03(2015)032} {\bibfield  {journal}
  {\bibinfo  {journal} {JHEP}\ }\textbf {\bibinfo {volume} {03}},\ \bibinfo
  {pages} {032} (\bibinfo {year} {2015})},\ \Eprint
  {http://arxiv.org/abs/1408.2831} {arXiv:1408.2831 [hep-th]} \BibitemShut
  {NoStop}%
\bibitem [{\citenamefont {Hristov}\ \emph
  {et~al.}(2019{\natexlab{b}})\citenamefont {Hristov}, \citenamefont
  {Katmadas},\ and\ \citenamefont {Toldo}}]{Hristov:2018spe}%
  \BibitemOpen
  \bibfield  {author} {\bibinfo {author} {\bibfnamefont {K.}~\bibnamefont
  {Hristov}}, \bibinfo {author} {\bibfnamefont {S.}~\bibnamefont {Katmadas}}, \
  and\ \bibinfo {author} {\bibfnamefont {C.}~\bibnamefont {Toldo}},\ }\href
  {\doibase 10.1007/JHEP01(2019)199} {\bibfield  {journal} {\bibinfo  {journal}
  {JHEP}\ }\textbf {\bibinfo {volume} {01}},\ \bibinfo {pages} {199} (\bibinfo
  {year} {2019}{\natexlab{b}})},\ \Eprint {http://arxiv.org/abs/1811.00292}
  {arXiv:1811.00292 [hep-th]} \BibitemShut {NoStop}%
\bibitem [{\citenamefont {Bobev}\ \emph {et~al.}(2020)\citenamefont {Bobev},
  \citenamefont {Charles},\ and\ \citenamefont {Min}}]{Bobev:2020pjk}%
  \BibitemOpen
  \bibfield  {author} {\bibinfo {author} {\bibfnamefont {N.}~\bibnamefont
  {Bobev}}, \bibinfo {author} {\bibfnamefont {A.~M.}\ \bibnamefont {Charles}},
  \ and\ \bibinfo {author} {\bibfnamefont {V.~S.}\ \bibnamefont {Min}},\ }\href
  {\doibase 10.1007/JHEP10(2020)073} {\bibfield  {journal} {\bibinfo  {journal}
  {JHEP}\ }\textbf {\bibinfo {volume} {10}},\ \bibinfo {pages} {073} (\bibinfo
  {year} {2020})},\ \Eprint {http://arxiv.org/abs/2006.01148} {arXiv:2006.01148
  [hep-th]} \BibitemShut {NoStop}%
\end{thebibliography}%

\end{document}